\documentclass[aps,pra,reprint,twocolumn,superscriptaddress]{revtex4-1}
\usepackage[T1]{fontenc}
\usepackage{graphicx}
\usepackage{xcolor}
\usepackage{hyperref}
\usepackage{multirow}
\usepackage{booktabs}
\usepackage{amsmath}
\usepackage{subcaption}
\usepackage{xr}
\usepackage{tikz}
\usepackage{tkz-graph}
\usetikzlibrary{calc,positioning,shadows.blur,decorations.pathreplacing,decorations.markings}

\definecolor{DARKMAGENTA}{HTML}{af2f40}
\definecolor{MAGENTA}{HTML}{DE707F}
\definecolor{ORANGE}{HTML}{E38851}
\definecolor{GREEN}{HTML}{8CD499}
\definecolor{GRAY}{HTML}{555555}

\hypersetup{
    colorlinks=true,
    linkcolor=DARKMAGENTA,
    citecolor=DARKMAGENTA,
    urlcolor=DARKMAGENTA,
}

\newcommand\define[1]{\textsf{#1}} 
\newcommand\code[1]{\texttt{#1}} 
\newcommand\name[1]{#1} 

\begin{document}

\title{AiiDA 1.0, a scalable computational infrastructure for automated reproducible workflows and data provenance}



\author{Sebastiaan P. Huber\normalfont\textsuperscript{\dag,}}
\email{sebastiaan.huber@epfl.ch}
\affiliation{National Centre for Computational Design and Discovery of Novel Materials (MARVEL), \'Ecole Polytechnique F\'ed\'erale de Lausanne, CH-1015 Lausanne, Switzerland}
\affiliation{Theory and Simulation of Materials (THEOS), Facult\'e des Sciences et Techniques de l'Ing\'enieur, \'Ecole Polytechnique F\'ed\'erale de Lausanne, CH-1015 Lausanne, Switzerland}
\author{Spyros Zoupanos}
\thanks{These authors contributed equally to this work.}
\affiliation{National Centre for Computational Design and Discovery of Novel Materials (MARVEL), \'Ecole Polytechnique F\'ed\'erale de Lausanne, CH-1015 Lausanne, Switzerland}
\affiliation{Theory and Simulation of Materials (THEOS), Facult\'e des Sciences et Techniques de l'Ing\'enieur, \'Ecole Polytechnique F\'ed\'erale de Lausanne, CH-1015 Lausanne, Switzerland}
\author{Martin Uhrin}
\affiliation{National Centre for Computational Design and Discovery of Novel Materials (MARVEL), \'Ecole Polytechnique F\'ed\'erale de Lausanne, CH-1015 Lausanne, Switzerland}
\affiliation{Theory and Simulation of Materials (THEOS), Facult\'e des Sciences et Techniques de l'Ing\'enieur, \'Ecole Polytechnique F\'ed\'erale de Lausanne, CH-1015 Lausanne, Switzerland}
\author{Leopold Talirz}
\affiliation{National Centre for Computational Design and Discovery of Novel Materials (MARVEL), \'Ecole Polytechnique F\'ed\'erale de Lausanne, CH-1015 Lausanne, Switzerland}
\affiliation{Theory and Simulation of Materials (THEOS), Facult\'e des Sciences et Techniques de l'Ing\'enieur, \'Ecole Polytechnique F\'ed\'erale de Lausanne, CH-1015 Lausanne, Switzerland}
\affiliation{Laboratory of Molecular Simulation (LSMO), Institut des Sciences et Ing\'enierie Chimiques, \'Ecole Polytechnique F\'ed\'erale de Lausanne (EPFL), Rue de l'Industrie 17, Sion, CH-1951 Valais, Switzerland}
\author{Leonid Kahle}
\author{Rico H\"{a}uselmann}
\affiliation{National Centre for Computational Design and Discovery of Novel Materials (MARVEL), \'Ecole Polytechnique F\'ed\'erale de Lausanne, CH-1015 Lausanne, Switzerland}
\affiliation{Theory and Simulation of Materials (THEOS), Facult\'e des Sciences et Techniques de l'Ing\'enieur, \'Ecole Polytechnique F\'ed\'erale de Lausanne, CH-1015 Lausanne, Switzerland}
\author{Dominik Gresch}
\affiliation{Microsoft Station Q, University of California, Santa Barbara, California 93106-6105, USA}
\author{Tiziano M\"{u}ller}
\affiliation{Institut f\"{u}r Physikalische Chemie, University of Z\"{u}rich, Switzerland}
\author{Aliaksandr V. Yakutovich}
\affiliation{National Centre for Computational Design and Discovery of Novel Materials (MARVEL), \'Ecole Polytechnique F\'ed\'erale de Lausanne, CH-1015 Lausanne, Switzerland}
\affiliation{Theory and Simulation of Materials (THEOS), Facult\'e des Sciences et Techniques de l'Ing\'enieur, \'Ecole Polytechnique F\'ed\'erale de Lausanne, CH-1015 Lausanne, Switzerland}
\affiliation{Laboratory of Molecular Simulation (LSMO), Institut des Sciences et Ing\'enierie Chimiques, \'Ecole Polytechnique F\'ed\'erale de Lausanne (EPFL), Rue de l'Industrie 17, Sion, CH-1951 Valais, Switzerland}
\author{Casper W. Andersen}
\author{Francisco F. Ramirez}
\author{Carl S. Adorf}
\author{Fernando Gargiulo}
\author{Snehal Kumbhar}
\author{Elsa Passaro}
\author{Conrad Johnston}
\affiliation{National Centre for Computational Design and Discovery of Novel Materials (MARVEL), \'Ecole Polytechnique F\'ed\'erale de Lausanne, CH-1015 Lausanne, Switzerland}
\affiliation{Theory and Simulation of Materials (THEOS), Facult\'e des Sciences et Techniques de l'Ing\'enieur, \'Ecole Polytechnique F\'ed\'erale de Lausanne, CH-1015 Lausanne, Switzerland}
\author{Andrius Merkys}
\affiliation{Vilnius University Institute of Biotechnology, Saul\.etekio al. 7, LT-10257 Vilnius, Lithuania}
\author{Andrea Cepellotti}
\author{Nicolas Mounet}
\author{Nicola Marzari}
\affiliation{National Centre for Computational Design and Discovery of Novel Materials (MARVEL), \'Ecole Polytechnique F\'ed\'erale de Lausanne, CH-1015 Lausanne, Switzerland}
\affiliation{Theory and Simulation of Materials (THEOS), Facult\'e des Sciences et Techniques de l'Ing\'enieur, \'Ecole Polytechnique F\'ed\'erale de Lausanne, CH-1015 Lausanne, Switzerland}
\author{Boris Kozinsky}
\affiliation{John A. Paulson School of Engineering and Applied Sciences, Harvard University, Cambridge, Massachusetts 02138, United States}
\affiliation{Robert Bosch LLC, Research and Technology Center North America, 255 Main St, Cambridge, Massachusetts 02142, USA}
\author{Giovanni Pizzi}
\email{giovanni.pizzi@epfl.ch}
\affiliation{National Centre for Computational Design and Discovery of Novel Materials (MARVEL), \'Ecole Polytechnique F\'ed\'erale de Lausanne, CH-1015 Lausanne, Switzerland}
\affiliation{Theory and Simulation of Materials (THEOS), Facult\'e des Sciences et Techniques de l'Ing\'enieur, \'Ecole Polytechnique F\'ed\'erale de Lausanne, CH-1015 Lausanne, Switzerland}

\date{\today}

\begin{abstract}
The ever-growing availability of computing power and the sustained development of advanced computational methods have contributed much to recent scientific progress.
These developments present new challenges driven by the sheer amount of calculations and data to manage.
Next-generation exascale supercomputers will harden these challenges, such that automated and scalable solutions become crucial.
In recent years, we have been developing AiiDA (\href{http://www.aiida.net}{aiida.net}), a robust open-source high-throughput infrastructure addressing the challenges arising from the needs of automated workflow management and data provenance recording.
Here, we introduce developments and capabilities required to reach sustained performance, with AiiDA supporting throughputs of tens of thousands processes/hour, while automatically preserving and storing the full data provenance in a relational database making it queryable and traversable, thus enabling high-performance data analytics.
AiiDA's workflow language provides advanced automation, error handling features and a flexible plugin model to allow interfacing with any simulation software.
The associated plugin registry enables seamless sharing of extensions, empowering a vibrant user community dedicated to making simulations more robust, user-friendly and reproducible.
\end{abstract}


\maketitle

\section*{Introduction}
Reproducibility is one of the cornerstones of the scientific method, as it enables the validation and verification of scientific findings~\cite{Ioannidis:2009,Peng:2011,Stoddart:2016,Allison:2016}.
In computational science, for a result to be reproducible, it should be possible to exactly retrace all the data transformations that led to its creation.
With the ever-growing availability of computational power, and increasingly complex computational workflows resulting in large amounts of interconnected data, \emph{a posteriori} reconstruction of provenance has become an intractable task.
Instead, to guarantee reproducibility, \emph{a priori} provenance should be effortlessly enforced through mechanisms that automatically track data \emph{as} it is being created.
Crucially, such a system should not merely store the data itself, but also preserve the explicit connection to the process that generated it, as well as the inputs of the latter and (recursively) their respective provenance.

While essential, automated provenance storage is not the sole requirement that must be satisfied to deliver effective reproducibility.
Rather, the data model should be generic enough to be applicable to any computational domain, and the infrastructure should be flexible enough to be interfaced with the diverse range of existing computational software.
Moreover, the infrastructure should also provide a system to fully automate complex simulations through the definition of robust workflows.
Lastly, data produced and stored should be easily shareable, such that it can be found, accessed and reused with different tools, following FAIR data principles~\cite{Wilkinson:2016}.
Such concepts of automated workflows, data management and sharing were formalised in the Automation, Data, Environment and Sharing (ADES) model and implemented in the AiiDA informatics infrastructure~\cite{Pizzi:2016}.

Hundreds of computational workflow management systems have been devised over the past decades~(\href{https://s.apache.org/existing-workflow-systems}{s.apache.org/existing-workflow-systems}).
One popular design choice is to use markup languages to describe static workflow logic.
While this allows to keep the workflow syntax simple, it tends to come at the cost of sacrificing the flexibility needed to dynamically change the execution path taken depending on the intermediate results (or errors) encountered.
However, this flexibility is crucial in the field of computational materials science~(see Ref.~\cite{Uhrin:2020} for an extensive discussion), which has been the main driver for the development of AiiDA so far.

GC3Pie~\cite{Maffioletti:2012}, Atomate~\cite{Mathew:2017}, Signac~\cite{Adorf:2018} and Parsl~\cite{Babuji:2019} are examples of workflow managers in the domain of computational science and high-performance computing, that have chosen to start from the Python programming language instead.
Like AiiDA, they provide new Python constructs to design and automatically execute workflow (in parallel), and store the resulting data.
The designs of these Python-based workflow managers differ depending on the typical use cases that they target; Parsl aims for many ($>10\,000$) short ($<100$ seconds) concurrent tasks, while Atomate and GC3Pie focus on fewer but computationally more demanding tasks, with Signac somewhere in between.
All frameworks store the data produced, but none with a focus on explicitly recording their provenance in detail, and in particular, storing the interconnections between data and the processes that received it as input or produced it as their output.
A focus on data provenance, as a core principle in AiiDA's design, is what currently differentiates it from the other workflow management systems mentioned above.

Early versions of AiiDA have already been successfully used in high-throughput computational studies~\cite{Mounet:2018,Kahle:2020,Mercado:2018,Prandini:2018,Vitale:2019}, making their AiiDA databases publicly available, and the corresponding provenance graphs interactively browsable through uploads on the Materials Cloud~\cite{Talirz:2020}.
As a consequence of such uptake of AiiDA, the size and scope of the needs quickly increased, testing the limits of the high-throughput capabilities of the original architectural design.
The workflow engine, the component responsible for the automated managing of all calculations and workflows, showed its first scalability limits under heavy computational loads when trying to manage hundreds of jobs concurrently on different computational resources.
AiiDA 1.0, released in October 2019, represents the culmination of a complete redesign that greatly improves efficiency and aims to deal with the high-throughput loads expected for upcoming exascale computing systems.
In addition to the improved engine, AiiDA 1.0 developed core new features that, amongst others, facilitate querying the provenance graphs and extend core functionalities, while remaining faithful to the principles of the ADES model.
Despite fairly radical changes to the source code, specific effort has been dedicated to guarantee that existing databases, generated with earlier versions, can automatically be migrated, guaranteeing and preserving the longevity of existing data.
In this paper, we first briefly describe the novel architecture of AiiDA 1.0, followed by a more in-depth discussion of new designs and features.

\section*{Architecture overview}
\begin{figure*}[ht!]
    \centering
    \includegraphics[width=0.8\linewidth]{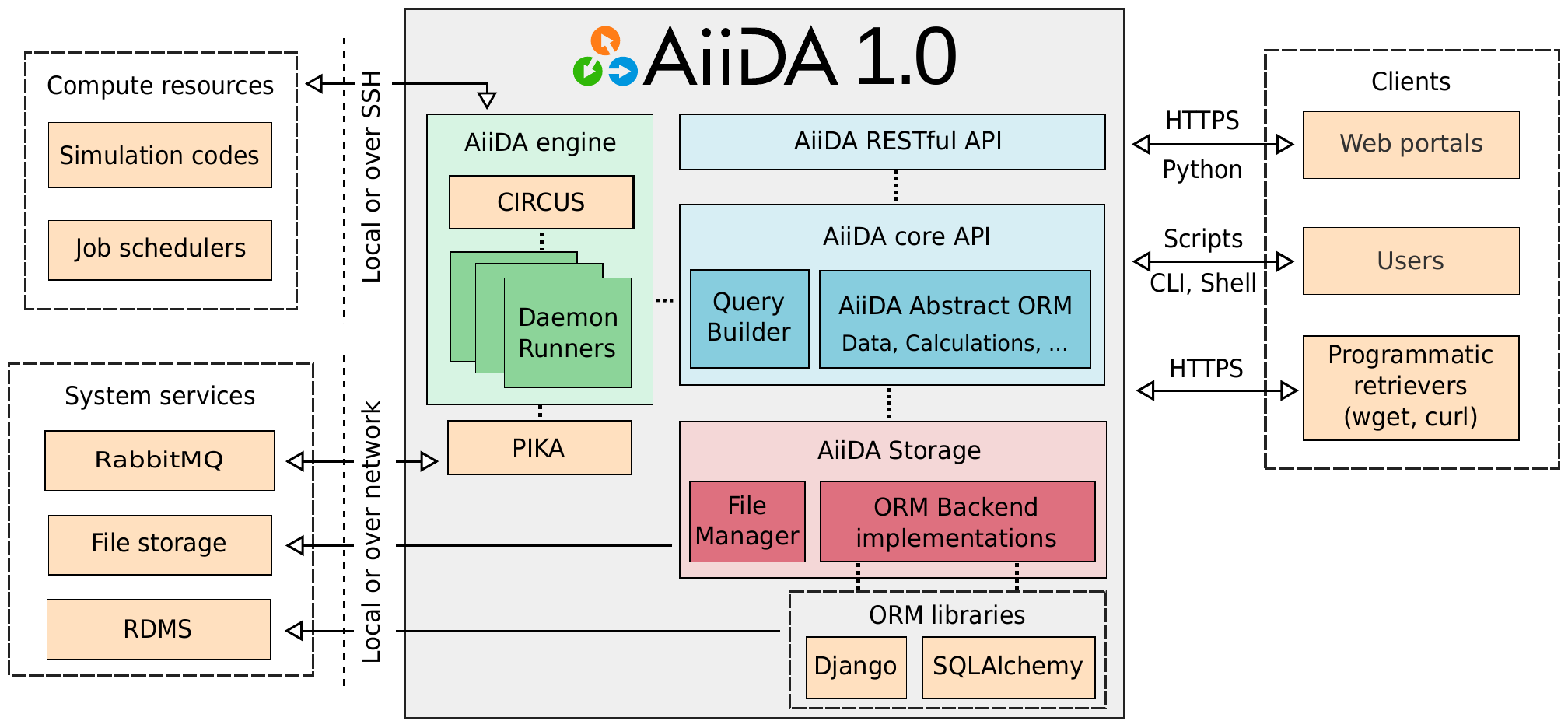}
    \caption{
        Schematic overview of the architecture of AiiDA 1.0.
    }
    \label{fig:architecture}
\end{figure*}

AiiDA aims to provide a framework that allows designing and running of high-throughput complex computational workflows with full automatic provenance and built-in support for high-performance computing on remote supercomputers.
The architecture, as shown in Fig.~\ref{fig:architecture}, is designed with these goals in mind.
A more detailed overview and a comparison with the architecture of earlier AiiDA versions can be found in section~A of the Supplementary Information.

One of the core components, the engine, is responsible for running all calculations and workflows that are submitted by the user and will described in detail in the section ``The engine''.
Calculations and workflows can be implemented in the custom language provided by AiiDA's core API which is implemented in Python.
An example of part of the workflow syntax is shown in Supplementary Fig. S2 and a more detailed description of the user interface can be found in Ref.~\cite{Uhrin:2020}.

Any calculation or workflow that is run by the engine will be automatically recorded in the provenance graph in order to enable the reproducibility of the results.
The definition and implementation of the provenance graph is explained in greater detail in section ``The provenance model''.
Besides the workflow language, the ORM also provides the tools to interact with the nodes of the provenance graph and inspect their content.
The QueryBuilder is the tool that allows efficient traversal of the provenance graph to select (sets of) nodes of interest and is described in detail in section ``Database abstraction and querying language''.

The contents of the provenance graph are stored in a file repository on the local file system and a relational database.
The mapping between the database and the Python API is performed by an Object Relational Mapper (ORM): currently the user can choose between the Django~(\href{https://www.djangoproject.com/}{djangoproject.com}) or SQLAlchemy~(\href{https://www.sqlalchemy.org/}{sqlalchemy.org}) library.

From the outside, users can interact with AiiDA through a command line interface called \code{verdi}, an interactive python shell or normal python scripts.
The REST API allows to query the provenance graph through HTTP calls (see section ``The REST API'' for more details).
AiiDA itself can communicate with computing resources either locally or over SSH to run calculations on those resources and comes with built-in support for most well-known and used job schedulers.

\section*{The engine}
\label{sec:engine}
The engine is the component of AiiDA in charge of automating the execution of calculations and workflows.
AiiDA is capable of managing calculations either on the local computer where AiiDA is installed, or on any number of remote resources (see section ``Running on external computers: calculation jobs'').
In addition, AiiDA provides a workflow language to define the logic to run complex sequences of steps, with potentially nested subworkflows and calculations.

The engine consists of runners that are executed in parallel as different operating system processes, supervised by a daemon that monitors and relaunches them if they were to die.
Each runner can process tasks independently and concurrently, distributing the workload involved in workflow and calculation execution.
Task distribution is achieved via a task queue implemented using the \name{AMQP} protocol through \name{RabbitMQ}~(\href{https://www.rabbitmq.com}{rabbitmq.com}), to guarantee reliable scheduling and almost instantaneous reaction to events such as the request for the submission of a new calculation or workflow, or continuing a workflow when the calculations or subworkflows it depends upon are completed.
Additional technical details on the implementation of the engine and the use of the event-based task queue can be found in the Methods section.

Thanks to this scalable architecture, the AiiDA engine is able to sustain high-throughput workloads involving tens of thousands of concurrent tasks every hour distributed on multiple computational resources, as we demonstrate in the Methods section ``Performance''.
Additionally, during execution AiiDA automatically stores all data and actions in the provenance graph (see ``The provenance model''), including the workflows, the calculations and their inputs and outputs, to provide full traceability.

In this section we define the three core concepts defined by the AiiDA engine (processes, calculations and workflows) and we briefly describe the key aspects of the engine implementation.

\subsection*{Processes in AiiDA}
In AiiDA, any entity that handles input data to produce output data, and that is run by the engine, is called a \define{process}.
Processes come in two flavours: \define{calculations} and \define{workflows}.
These two terms in AiiDA have a more specific meaning than their use in common parlance.
In particular, calculations are defined as processes that create new data as output, given certain data as input.
A typical case is the execution of a simulation code on a remote computer.
In contrast, workflows in AiiDA are solely tasked with the orchestration of subprocesses, calling calculations and/or other workflows in a certain logical sequence.
Consequently, workflows are not allowed to generate new data, but can only return existing data, usually created by one of the calculations that they called (either directly or indirectly via a subworkflow).

This distinction is critical in the design of the provenance model of AiiDA, allowing to distinguish the part of the provenance graph that represents exactly how data was generated (the \define{data provenance}) from the \define{logical provenance} that captures the why of the data creation, i.e., which workflows drove the execution.
This distinction is elaborated in more detail in the subsection ``Logical and data provenance''.

In routine tasks, users do not interact directly with calculations or workflows, but with specific subclasses that define additional functionalities and are appropriate in different use cases that we describe below.

\subsection*{Process functions: calculation functions and work functions}
The simplest way to define a process (a calculation or a workflow) is to add a line (\code{@calcfunction} or \code{@workfunction}, respectively) on top of a standard python function (this line is called a decorator in Python; see the supplementary section ``Example of a work chain and calculation function'' for an example).
Calculation and work functions are collectively referred to as process functions.
The logic of the \code{@calcfunction} and \code{@workfunction} decorators is defined by AiiDA, and they signal to AiiDA's engine that any time the function is called, its execution should be recorded in the provenance graph, linking up the inputs passed and the outputs it produced.
Furthermore, if a work function calls another process function, this action is also represented in the provenance graph as a connection between the caller and the called process.

This approach of defining processes is very intuitive and powerful, because any standard Python function can be converted into a process function simply by adding a decorator, as long as it accepts AiiDA datatypes as input and returns an AiiDA datatype (or a dictionary of AiiDA datatypes) as output.
However, process functions show their limits when running very long workflows that can last for days (e.g., in the case of molecular-dynamics simulations).
The reason is that process functions are executed on the local machine and since it is not possible to interrupt the execution of a Python function and resume it later, the local machine cannot be shut down, rebooted or disconnected from the network until the process is finished.
In addition, the interface design, through its simplicity, restricts what code can be executed, as it has to be written in Python.
Of course one could resort to calling external codes through subprocesses, but this quickly becomes complicated, especially if the external code needs to be run on a remote machine and/or through a job scheduler.
For these reasons, AiiDA implements two additional processes: \define{calculation jobs}, to manage the execution of external codes through job schedulers, and \define{work chains} to implement and orchestrate long-running workflows with the possibility of pausing and restarting between steps.

\subsection*{Running on external computers: calculation jobs}
\label{sec:calculation_jobs}
Calculations jobs, implemented by the \code{CalcJob} process class, are used to manage the execution of external codes, commonly run via a job scheduler and optionally on a remote machine.
The \code{CalcJob} can be adapted for any external code through plugins (see ``The plugin system''), which define how the required raw input files are constructed from the AiiDA datatypes that are provided as input.
Once submitted, the engine takes over and performs all the necessary steps to run the calculation to completion.
These include uploading the raw input files, submitting the job to the scheduler, querying the job state and waiting for it to finish, and finally retrieving the files to be stored locally.
The retrieved files can optionally be parsed into AiiDA datatypes that are registered as outputs of the calculation.
The parser logic is also defined through plugins, since this also requires code-specific logic.
The explicit parsing into AiiDA datatypes makes the outputs interoperable among different codes and allows for their direct reuse as inputs of new calculations.

To increase the robustness of the engine, we implemented optimised algorithms to automatically reschedule failed tasks through an exponential-backoff mechanism, recovering from common transient errors such as connection issues.
If the task fails multiple times consecutively (for example because the remote machine went offline), the process is automatically paused.
The process can be resumed effortlessly by the user when the issue is resolved, and the AiiDA engine ensures that no data loss occurs as the calculation resumes from where it was paused.

Additionally, to avoid overloading remote machines with connections, rendering them potentially unresponsive also to other users, AiiDA implements a connection pooling algorithm.
Connections to the same computer are grouped and new requests are funnelled within existing open connections, if available.
If no connection is open and a new one needs to be created, AiiDA guarantees that a configurable minimal time interval between them is respected.
For similar reasons, AiiDA caches the list of jobs in the scheduler queue and respects a minimum (configurable) timeout when refreshing their state.

\subsection*{Interruptible workflows: work chains}
As discussed earlier, the limitation of a workfunction is that its execution is atomic and blocking.
Therefore, if workfunctions are used to implement complex workflows calling many long-running calculation jobs, intermediate progress within the function body cannot be persisted.
As a consequence, if the process is interrupted (for instance to restart the AiiDA daemon or to reboot the machine where AiiDA is running), all progress is lost.

To solve this issue, AiiDA implements the \code{WorkChain} process class.
Work chains allow users to specify units of work (like the processing of data or the submission of subprocesses) as steps.
The logical outline of the steps (i.e., their sequence, possibly including \code{while} and \code{if/else} logic) is defined in the process specification.
By design, the execution of the workflow is managed by the AiiDA engine that gets back control between steps, executing the next one only when all subprocesses have finished.
Most importantly, between steps the engine persists the workflow state to the database.
Therefore, if the engine is stopped no progress is lost and, upon restarting, the engine will continue the workflow execution from the last persisted step (or the engine will keep waiting for running subprocesses, if any).

In addition to the capability of stopping and continuing execution, work chains have the advantage that inputs and outputs (including information like their type or whether they are required or optional) can be declared in the process specification, together with the logical outline of the steps.
Work chains are thus self-documenting because, through simple inspection of the specification, users can determine the interface of the workflow and its main logic.
Furthermore, the specification is machine-readable and can be automatically rendered in multiple formats.
One use case is the extension provided in AiiDA for the Sphinx~(\href{http://sphinx-doc.org}{sphinx-doc.org}) engine (used to generate the AiiDA documentation, see ``Community building'') that generates human-readable documentation for any work chain, to be displayed for instance as a web page.

The work chain interface encourages writing modular workflows, with lower-level work chains implemented to solve specific tasks that are often code-dependent, like error handling, restarting or automatic tuning of parameters.
Higher-level work chains can then wrap around these, directly exposing all inputs of the lower-level workflows or only a few, with the others being determined by the workflow logic.
Such wrapping work chains can implement additional functionality less focused on the technical details of the code execution but rather focusing on the evaluation of scientific quantities of interest.
As such, workflows enable the delivery of fully automated turn-key solutions that require only minimal inputs and encode the scientists' knowledge on how to compute a given result, from code-specific technicalities to the handling of the parameters involved in the scientific models of the simulations.

To conclude, we emphasise that implementing workflows directly in Python is a design choice.
Since there is no translation between what the user implements and what the engine executes, the user can leverage the power of the full Python and AiiDA APIs when designing workflows.
Additionally, this design directly facilitates debugging using standard Python tools and seamless integration with external libraries for data manipulation.

\section*{The provenance model}
\label{sec:provenance}
\begin{figure}[tbp]
    \centering
    \includegraphics[width=0.48\textwidth]{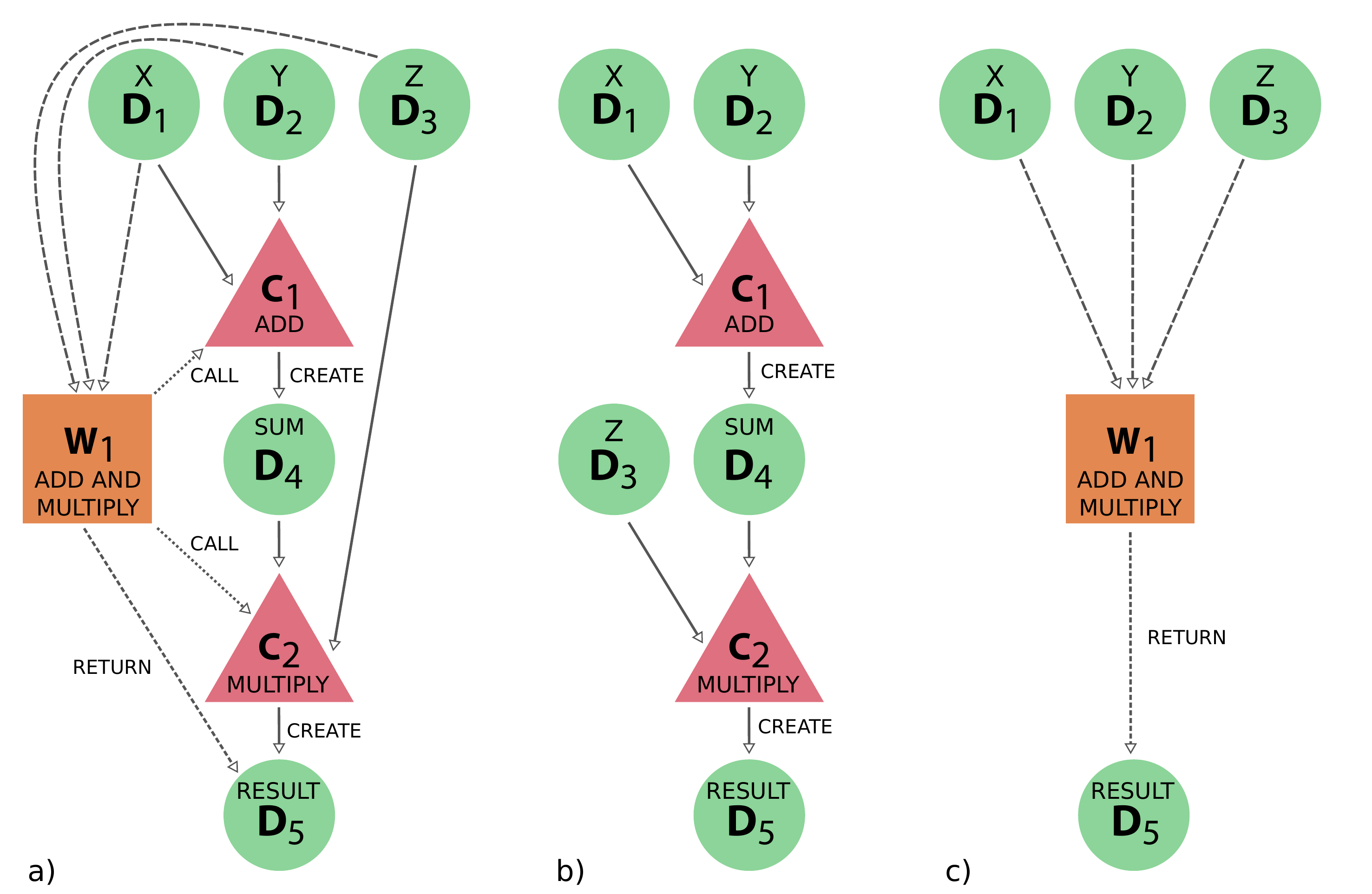}
    \caption{
        (a) A schematic provenance graph representing the execution of a workflow $W_1$ receiving three data nodes $D_1$, $D_2$ and $D_3$ as input, containing the values $x$, $y$ and $z$ respectively. $W_1$ computes the expression $(x+y)\cdot z$ by calling two calculations $C_1$ (to perform the sum) and $C2$ (to perform the product), forwarding the correct inputs to them. $C_1$ creates the intermediate node $D_4$ (with the value $x+y$) and $C_2$ then creates the node $D_5$ with the final result, that is then also returned by $W_1$.
        While this simplified example is purely for illustrative purposes, it demonstrates that by storing execution information as a graph, the provenance of all data is fully recorded.
        (b) Data-provenance layer: it includes calculation and data nodes only, showing the exact sequence of steps that led to the creation of the data nodes.
        (c) Logical-provenance layer: it hides the details of all intermediate results and focuses only on how the workflow produced the final results from a given set of inputs.
    }
    \label{fig:provenance_graph}
\end{figure}

As explained in the previous section, AiiDA's engine automatically represents the execution of processes, along with their inputs and outputs, as vertices (called \define{nodes} in AiiDA) of a directed graph.
Graph edges (called \define{links}) connect the nodes; process nodes, for instance, have incoming links from their inputs and outgoing links to their outputs.
Since output data can in turn be used as input to new processes, extensive graphs are generated.
We call them \define{AiiDA provenance graphs} because they allow to retrace the exact steps that led to the creation of a piece of data.
An example of a simple provenance graph is shown in Fig.~\ref{fig:provenance_graph}.

\subsection*{Node types}
\label{sec:node_link_types}
While all nodes of the graph share a set of common properties, there is a need to define custom properties based on what the node represents.
Therefore, the various AiiDA processes (see ``The engine'') as well as data are represented by different node subtypes.
This makes it possible to implement functionality specific to each of them and to explicitly target nodes of a certain type when querying the graph (see ``Database abstraction and querying language'').

\begin{figure*}[tbp]
    \centering\includegraphics[width=.98\linewidth]{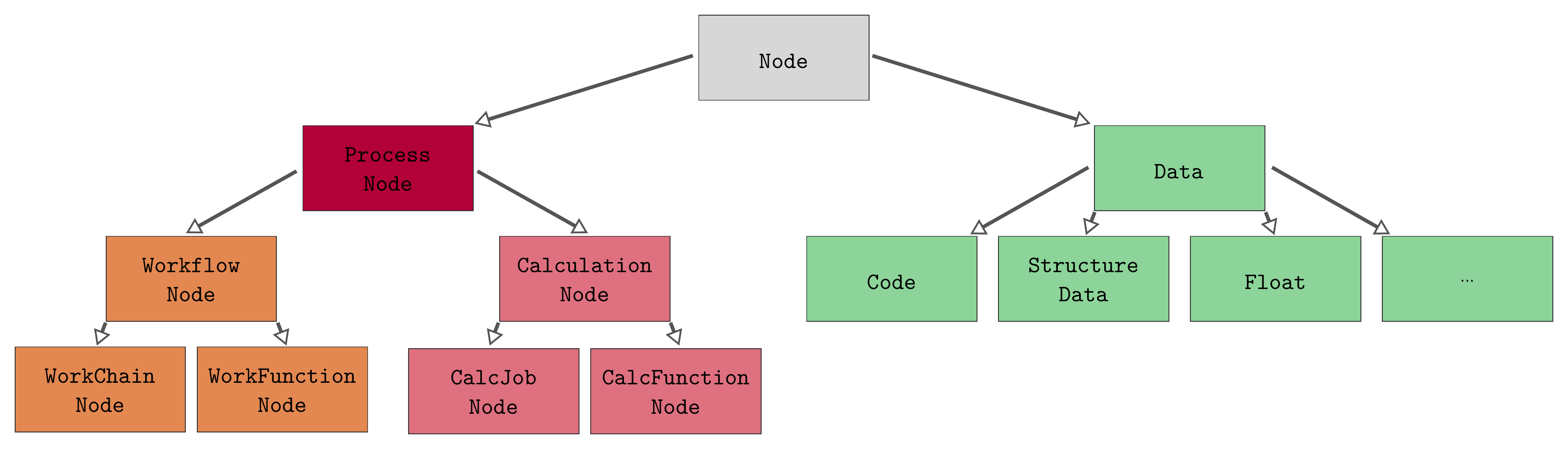}
    \caption{
        The hierarchy of the node types in AiiDA.
        This hierarchy is also mirrored in the Python code, where Python classes are used to represent them, using Python's inheritance model.
        The different node classes allow to implement custom functionality for each subtype.
        Additionally, the subclass hierarchy allows to query for specific node types, or a set thereof.
    }
    \label{fig:node_hierarchy}
\end{figure*}

In AiiDA, the \code{Node} class is the base class to represent any node in the graph.
The common properties of any node include the user who created it, the creation and last modification times, an optional computer on which it was run or stored, and a human-readable label and description.
\code{Node} classes are subclassed to build a hierarchy of node types, schematically represented in Fig.~\ref{fig:node_hierarchy}.
In particular, data and process nodes are represented by the \code{Data} and \code{ProcessNode} subclasses, respectively.

Various data types are implemented by directly subclassing the \code{Data} class.
AiiDA ships with a few basic data types; for instance, among many others, \code{Float} to represent a single float value and \code{Dict} for a dictionary of key-value pairs.
Arbitrary new data types can be defined through the plugin system (see ``The plugin system'').

The hierarchy of \code{ProcessNode} subclasses reflects the distinction in AiiDA between calculations and workflows, represented by subclasses of \code{CalculationNode} or \code{WorkflowNode}, respectively.
In practice, in a provenance graph one finds instances of \code{CalcJobNode}, \code{CalcFunctionNode}, \code{WorkChainNode} or \code{WorkFunctionNode}, representing executions by the engine of the corresponding process classes (\code{CalcJob}, \code{calcfunction}, \code{WorkChain} or \code{workfunction}, respectively, as discussed in ``The engine'').
The intermediate classes in the hierarchy serve mostly as a taxonomic classifier and they are useful when querying the provenance graph (see ``Database abstraction and querying language'').
For instance, querying for \code{WorkflowNode}s will match both \code{WorkChainNode}s as well as \code{WorkFunctionNode}s.

\subsection*{Link types}
\begin{figure}[tbp]
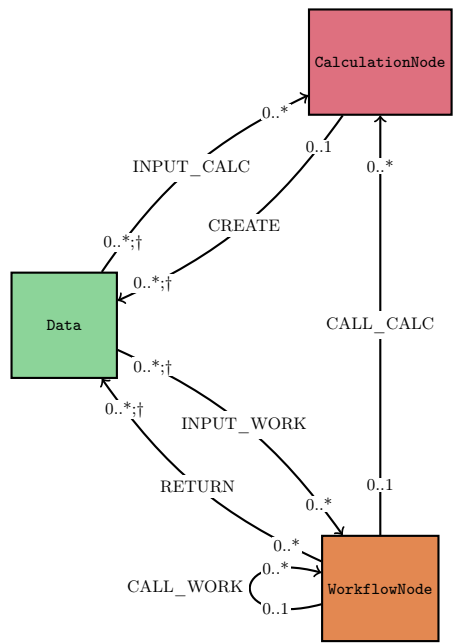

    \centering
    \include{figures/orm_main_and_sub_relationships}
    \caption{
        Link types allowed in the AiiDA provenance graph.
        Rectangles represent node types and arrows connecting them indicate the direction and the type of each link.
        The symbols at the start and end of each arrow indicate the cardinality of the corresponding link types: \code{0..1} means that at most one node is allowed on that link endpoint for a given node on the opposite endpoint (for instance, a \code{Data} node can have at most one \code{CalculationNode} as its creator); \code{0..*} means that any number of nodes is possible (for instance, a \code{CalculationNode} can have an arbitrary number of input \code{Data} nodes).
        Additionally, a dagger ($\dagger$) indicates that link labels must be unique for a given node on the opposite endpoint (for instance, outgoing create links from a \code{CalculationNode} must have unique labels).
    }
    \label{fig:links}
\end{figure}

All links have a type to indicate the semantic meaning of the relationship.
In addition, links have a label that can be used, given a node, to distinguish nodes connected to it with the same link type.
For example, labels identify the different input nodes to a process.
A summary of all link types in AiiDA is shown in Fig.~\ref{fig:links}.

Process nodes can have input and output links to data nodes, representing their inputs and outputs. More specifically, in the implementation, input links can be of type \code{INPUT\_CALC} and \code{INPUT\_WORK} depending on the type of the linked process node.
Similarly, output links can either be of type \code{CREATE} or \code{RETURN} (for calculation and workflow nodes, respectively) explicitly highlighting the difference between calculation and workflow processes described in ``The engine''.

In addition to links between data and process nodes, two additional link types exist between processes, to indicate that a workflow called another process.
These links are called \code{CALL\_CALC} and \code{CALL\_WORK}, depending on the type of the called process, and are collectively referred to as call links.

Due to their semantic meaning, link types impose precise validation rules.
First, a link type requires specific node types at its two endpoints.
In addition, cardinality rules are defined, as illustrated in Fig.~\ref{fig:links}.
In general these dictate that, given a node, any number of nodes can be linked to it by links of a given type.
For create and call links there are instead explicit restrictions: a data node can be created by at most one calculation, while it can be returned by multiple workflows; a process can be called by at most one workflow, while a workflow can call an arbitrary number of subprocesses.
Finally, uniqueness constraints on the labels of certain link types are enforced: the link labels of input nodes to a given process must be unique, to guarantee that each input can be uniquely identified.
The same applies to the output nodes of a process.

\subsection*{Logical and data provenance}
\label{sec:provenance_logical_data}
Due to the rules defined in the previous sections, AiiDA provenance graphs have useful properties.
The subgraph composed exclusively of data and calculation nodes together with the links that connect them forms a \define{Directed Acyclic Graph} (DAG).
The acyclicity is a direct result of calculations only having outgoing links to data they created: output nodes are always generated as a result of execution and therefore, because of the causality principle, cannot also simultaneously be part of the inputs of the process itself or of any parent process in the graph.
We refer to this subgraph as the \define{data provenance}, since it is an exact record of the origin of the data.
Since a DAG has well defined properties, assumptions can be made by the AiiDA query language (see ``Database abstraction and querying language''), for instance to define the concept of an \define{ancestor} of a given node as any node that can be reached following the links of the data-provenance DAG in the backward direction (i.e., moving from the node at the head of a link to the one at the tail).
Similarly, \define{descendants} are defined as nodes that can be reached following DAG links in the forward direction.

We can also define a second subgraph, composed solely of data and workflow nodes and the links connecting them (including \code{CALL\_WORK} links).
We call this graph the \define{logical provenance}, since it focuses on the logical steps taken by the workflows when processing data and orchestrating processes.
Like the data provenance, the logical provenance subgraph also forms a directed graph that, however, is not acyclic.
In fact, return links are not bound by the causality principle since they do not signify ``creation'', and workflows merely return outputs that have already been created by other calculations.
For instance, a workflow selecting one of its inputs based on some criteria will have a return link to that input, which introduces a cycle in the graph.

The two layers of data and logical provenance share data nodes and are interconnected by \code{CALL\_CALC} links.
The separation of these two provenance layers has the additional benefit of allowing to select the granularity for the inspection of the provenance graph, as shown in Fig.~\ref{fig:provenance_graph}(b)-(c).

\subsection*{Node properties}
\label{sec:provenance_node_properties}
In the previous section ``Link types'', we described the basic properties that are common to all node types.
The various subtypes define the ``schema'' of the additional properties that fully define and describe the node.
AiiDA provides two data stores to persist these properties: a filesystem repository and a relational database (where any JSON-serialisable key-value pair can be saved, see ``Database abstraction and querying language'').
Properties that are stored in the database are named \define{attributes}, which are fully and efficiently queryable.
In contrast, properties that do not require querying and/or are very large in size, such as large arrays or raw files, are better stored in the repository so as not to overburden the database.
Attributes and the files in the repository are immutable once the node is stored, since together they define the ``content'' of the node and allowing them to be changed would invalidate the provenance of descendants.
Mutable properties are also allowed and are called \define{extras} that, like attributes, are stored as key-value pairs in the database.
However, in stark contrast to attributes, extras can be added and/or modified at any time.
A typical use case is to tag nodes with custom properties that can, for example, be used for more selective querying.

\subsection*{Reproducibility and efficiency}
By automatically recording all data transformations through a process together with all the inputs it consumed, the produced graph is in principle fully reproducible.
This of course depends on all relevant inputs being stored as well as the code that was ran by the process.
The actual source code of process functions is stored by AiiDA in the repository; for all other processes the code is referenced indirectly by storing both the version of AiiDA and of the relevant plugin in the attributes of the process node.

Tracking all data transformations in a provenance graph provides another benefit besides enabling reproducibility.
The entire provenance graph can serve to implement a ``caching'' mechanism: if one considers the execution of a process that already has been performed before with identical inputs, the actual execution can be skipped.
In this case, since the inputs are identical, we already know what the outputs are going to be as well, and so we can simply take those from the previously executed process instead, avoiding to incur the computational cost once more.
This caching mechanism is implemented in AiiDA 1.0 and is explained in detail in the section ``Caching''.

\section*{Database abstraction and querying language}
\label{sec:queryinglanguage}
\begin{figure*}[t!]
    \centering
    \includegraphics[width=0.7\linewidth]{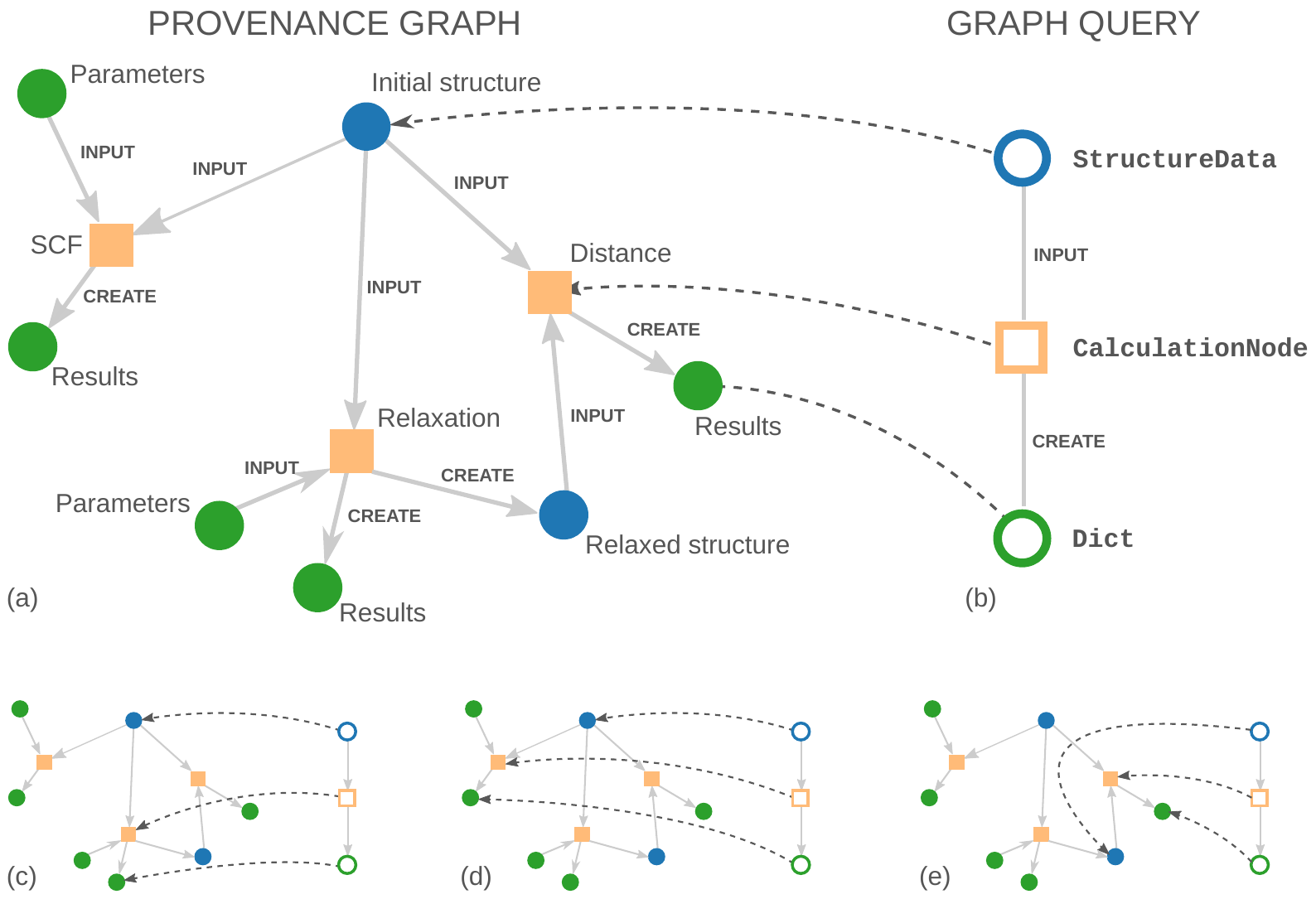}

    \caption{
        \label{fig:queryembeddings}
        (a) Schematic of an AiiDA graph that could result from a materials science simulation: as described by the labels, a Density Functional Theory self-consistent field (SCF) calculation and a geometry relaxation of a crystal structure, and a calculation of the ``distance'' between the initial and final structure.
        Orange squares represent nodes of type \code{CalculationNode}, circles represent \code{Data} nodes: blue for crystal structures (of type \code{StructureData}) and green for nodes of type \code{Dict} (dictionaries of key--value pairs with input parameters or parsed results).
        (b) Representation of a graph query searching a \code{StructureData} node that was an input of a \code{CalculationNode} that created a \code{Dict} node as output.
        Labels on the right represent the filter on the node type applied while querying.
        One of the possible embeddings of this query in the graph is also shown between panels (a) and (b).
        (c, d, e) These panels show all other subgraphs that match the query embedded in the entire provenance graph.
    }
\end{figure*}

When running automated high-throughput research projects with AiiDA, very large provenance graphs containing millions of nodes or more are easily generated (as for instance in the study of Ref.~\cite{Mounet:2018}).
Tools that can efficiently query such graphs become essential to perform data analysis.
In order to allow the implementation of a performant tool, AiiDA uses a relational database to store the provenance graph with its links and nodes including most of their properties (see ``Node properties'').
In addition, we have optimised AiiDA's database schema and indexes to ensure that typical graph queries are very efficient, as discussed in section ``Performance''.

Direct database queries must be expressed in its native language.
AiiDA's current database solution \name{PostgreSQL}~(\href{http://www.postgresql.org/}{postgresql.org}) is based on the SQL language, which, while known for its efficiency, requires (in the context of the AiiDA's provenance graph) writing queries that are long and cumbersome even for database experts.
Furthermore, the exact query structure depends on the specific implementation choices of AiiDA, that could change between versions to improve efficiency (for instance if the schema is improved or the SQL backend is replaced by a graph database).
AiiDA does not directly write the SQL statements itself but relies on object-relational mapping (ORM) libraries like \name{Django} and \name{SQLAlchemy} as an intermediate layer to express the queries in Python.
However, these queries still depend on the specific database schema and have essentially the same complexity level as the native ones.
It is therefore crucial to provide a tool that abstracts this process and makes writing queries not just as simple as possible but also independent of the ORM and database implementation.

The \define{query builder} is the tool in AiiDA that satisfies these criteria: it allows users to express any query on the provenance graph using a familiar Python syntax that is automatically translated into an optimised SQL query and executed.
To illustrate the concept of the query builder, we describe in Fig.~\ref{fig:queryembeddings} a simple provenance graph and how a query is mapped onto it.
Essentially, this is the problem of subgraph isomorphism, which, given two graphs $G$ and $H$, consists of finding all the subgraphs within $G$ that are isomorphic to $H$~\cite{Ullmann:1976}.
Here $G$ is the entire AiiDA provenance graph and $H$ is the subgraph represented by the query as expressed through the query builder.

The schematic query shown in Fig.~\ref{fig:queryembeddings}(b) represents the search for all crystal structures (\code{StructureData}) used as input to a calculation (\code{CalculationNode}) that created a \code{Dict} node as output.
The query encodes filters on the link directions, link types (\code{INPUT}, \code{CREATE}) and node types (\code{StructureData}, \code{CalculationNode}, \code{Dict}).
On top of these constraints, additional query specifications are available which are not shown in Fig.~\ref{fig:queryembeddings} but are explained in detail in the ``Query builder syntax example'' section in the Supplementary Information.
In particular, filters can be set on the node properties, for example the structure must contain a given chemical element or the output must have a value within a certain range.
Moreover, the user can specify a list of \define{projections}, i.e., which subset of properties of the matched nodes should be returned.
Once the query is fully defined and executed by the user, the query builder returns all the subgraphs embedded in the AiiDA provenance graph that match the query constraints.
Fig.~\ref{fig:queryembeddings} shows all four embedded subgraphs that match the query of this particular example.
Finally, the query builder converts the results into Python objects, that can directly be used with common data processing libraries like \name{SciPy}~(\href{https://www.scipy.org/}{scipy.org}) and \name{pandas}~(\href{https://pandas.pydata.org/}{pandas.pydata.org}) for further data analysis.

The reason for the existence of two ORM backend implementations (Django and SQLAlchemy) is mostly historical.
The original implementation used only Django, and as a result AiiDA's API was tightly coupled to this library.
When the query builder was first introduced, interaction with the database was implemented instead via SQLAlchemy, as it provided a richer feature set allowing for more general queries.
Since SQLAlchemy at the time already provided support for JSONB (unlike Django), it was also used as an alternative ORM backend implementation to benefit from the significant improvements in database performance (see ``Performance'').
To achieve this, extensive work was performed to decouple the backend ORM from Django-specific constructs, and to create a new layer of backend-independent AiiDA \define{frontend classes} (those directly exposed to users to interact with nodes).
Moreover, all functionalities of AiiDA (command line tools, query builder, import/export functionality, \ldots) were critically revised and updated so as not to rely anymore on backend-specific logic but to use general backend-independent interfaces instead.
As a consequence, in the current version of AiiDA the user interface does not depend anymore on the chosen backend.
Additionally, implementing new ORM backends is now greatly streamlined and simplified.
Django also started supporting JSONB recently and, therefore, we have updated the Django interface in AiiDA to use JSONB fields for attributes and extras instead of our original solution based on entity-attribute-value tables, increasing database performance also for the Django backend (see ``Performance'').

\section*{The REST API}
\label{sec:restapi}
The query builder is the tool of choice for querying data directly via the Python interface, which in turn is the preferred approach when one has direct access to the machine running AiiDA.
If, instead, data needs to be made available to users without full access to the machine, for example over the web, a solution is required that can serve the data in a secure way.
In addition, such a solution should not just be able to serve the database contents as a whole, but it should provide the necessary functionality to query for specific data.

A widespread approach to share data over the web is through a REST API~(\href{https://www.w3.org/2001/sw/wiki/REST}{w3.org/2001/sw/wiki/REST}), a stateless protocol that allows to query and retrieve data via HTTP requests~(\href{https://www.w3.org/Protocols/}{w3.org/Protocols}).
REST APIs are not only generally adopted on the web, but they have also become widespread in specific scientific disciplines and domains.
For instance, in the materials science community the OPTiMaDe~(\href{http://www.optimade.org}{optimade.org}) consortium has been formed to define a common REST interface for querying material property databases.
Many of the major materials databases are planning to implement the OPTiMaDe interface like AFLOW~\cite{Curtarolo:2012}, the Crystallography Open Database~\cite{Grazulis:2012}, the High-Throughput Toolkit~(\href{http://httk.openmaterialsdb.se}{httk.openmaterialsdb.se}), Materials Cloud~\cite{Talirz:2020}, Materials Project~\cite{Jain:2013}, Nomad~(\href{http://www.nomad-repository.eu/}{nomad-repository.eu}), OQMD~\cite{Kirklin:2015}, and AiiDA~(\href{http://www.aiida.net/}{aiida.net}), among others.

AiiDA implements a REST API server that can be launched directly from the command line or deployed via scalable web servers like Apache~(\href{https://httpd.apache.org}{httpd.apache.org}).
API endpoints are available to access data associated with the AiiDA graph, including the list of nodes, their properties (like attributes, extras and files in the repository) as well as the graph information (incoming and outgoing links for a given node).
Nodes are identified by their UUID~(\href{http://www.ietf.org/rfc/rfc4122.txt}{ietf.org/rfc/rfc4122.txt}) to ensure that resources are uniquely identifiable even if data is shared and then made available by a different server.
Custom queries can be performed by specifying filters in the query string in order to narrow the matched subset of nodes.
The web server is implemented using the \name{flask}~(\href{https://palletsprojects.com/p/flask/}{palletsprojects.com/p/flask}) web framework and a number of flask plugins (including \name{flask-sqlalchemy} to manage sessions to the database and \name{flask-restful} to handle requests using the REST approach).
Once a request is received, this is translated into a database query using AiiDA's query builder, which is then executed.
The results of the query are then mapped to the format defined by the REST API and serialised into a JSON response that is returned to the web client.
Results are paginated to facilitate downloads of large amounts of data without overloading the server.
In addition to endpoints common to all node types, additional endpoints are available that provide functionality specific to only certain node subtypes.
For example, it is possible to directly obtain raw inputs and outputs of a calculation, or to download the content of data nodes in a specific format other than AiiDA's internal representation.

In conclusion, AiiDA's REST API endpoints provide a complete set of features to interact with AiiDA programmatically and in a secure way from the web.
An emblematic example of its use is provided by the Explore section of the Materials Cloud~\cite{Talirz:2020} portal, where AiiDA databases are made available as interactive web pages and the provenance graph can be browsed via a graphical user interface.
Indeed, all data needed to display the provenance graph and the node contents on Materials Cloud Explore is obtained through AiiDA's REST API.

\section*{The plugin system}
\label{sec:pluginsystem}
The AiiDA ontology and provenance graph are designed to be employed in any field of computational science, and therefore they are intentionally domain agnostic.
To enable users to extend core functionalities to suit the needs of a specific discipline, AiiDA provides a flexible plugin system to add custom data types, to interface any simulation code with specific input generators and parsers, to implement custom workflows, and more.
These extensions are registered upon installation through entry points, as explained in ``Architecture'', which allows them to be developed and installed completely independently of the AiiDA code base.
To promote sharing of plugins, AiiDA provides an online registry (see ``Registry'') where plugin packages can be registered, discovered and downloaded.

\subsection*{Architecture}
\label{sec:plugin-system:architecture}
The plugin system builds on top of the \name{setuptools} project~(\href{https://setuptools.readthedocs.io/en/latest/}{setuptools.readthedocs.io}), currently the \emph{de-facto} standard tool for bundling and installing Python packages.
Setuptools provides a feature called ``entry points'', which are handles to specific Python resources (e.g., a class or function) of a package.
A Python package can define these entry points (categorised in entry point groups) such that, once installed, the corresponding resources are registered and become automatically discoverable and usable by any other package.
AiiDA leverages the entry point system by defining several groups specific to the type of resources that a plugin can extend, for example \code{aiida.data} and \code{aiida.workflows} for new data types and workflows, respectively (ten groups are currently defined, to extend support to new codes, parsers, job schedulers, transport protocols to connect to remote computers, external databases, etc.).
When AiiDA plugin packages register plugins in the appropriate entry point groups, AiiDA automatically discovers them and makes the functionality available to users, integrating it in its core infrastructure.
Plugin packages are encouraged to namespace their entry points by prefixing them with the package name to avoid overlap with entry points of other packages.

Installing a plugin package can be performed with a single command of the \define{pip}~(\href{https://pypi.org/project/pip/}{pypi.org/project/pip}) Python package manager if the package is published on \define{PyPI}~(\href{https://pypi.org}{pypi.org}).
We emphasise, however, that publication on PyPI is not required.
Local or private packages can be installed just as easily and they can also register plugin entry points, giving developers full freedom on how to maintain and distribute their plugin packages.

\subsection*{Registry}
\label{sec:plugin-system:registry}
The AiiDA plugin registry~(\href{https://aiidateam.github.io/aiida-registry//}{aiidateam.github.io/aiida-registry/}) is an online resource with a list of all known plugin packages.
This centralised overview makes plugins easily discoverable by users of the AiiDA community and encourages code sharing and reuse.
Authors can register their plugin packages through a pull request to the registry repository providing minimal information like package name, development status, links to the code repository and documentation, as well as a URI pointing to a JSON file maintained by the plugin developers.
The latter contains additional information in setuptools format, such as the name and description of the project, the authors list and the entry points that are provided.
Based on the information provided by the registered packages, the registry website is automatically built through continuous deployment.
The plugin system in combination with the registry provides a powerful and effective tool to allow users and developers to easily create and share extensions of AiiDA's core functionality.

\section*{Community building}
\label{sec:communitybuilding}
Science is a collective enterprise and scientific reproducibility can only be realised through a concerted effort of the scientific community.
Making a software tool like AiiDA available is an important first step towards improving the reproducibility of computational science, but it is not enough.
One needs to facilitate the uptake of the tool by the community, such that the data produced by it become interoperable and reusable.
We have undertaken various approaches to build a community around AiiDA, such as guaranteeing the quality and robustness of the code, ensuring the longevity of data produced with AiiDA, and actively promoting knowledge transfer by online and offline training of users and developers through tutorials and workshops.
Additionally, we are now member of NumFOCUS~(\href{https://numfocus.org/}{numfocus.org}), an organisation that promotes open science and open research data.

\subsection*{Code quality, testing and continuous integration}
AiiDA's source code is hosted on GitHub (\href{http://www.github.com}{github.com}) and all code development contributions, both from internal and external contributors, go through its pull request system.
These pull requests facilitate the review and improvement of the suggested changes before they are accepted into the main code base.
On top of this quality control performed by peers, we have enabled an automatic continuous integration testing system~\cite{Duvall:2007}.
Using GitHub actions (\href{https://github.com/features/actions}{github.com/features/actions}), each code commit triggers the running of a suite of over 1000 unit and integration tests that verify that the changes do not break existing functionality.
These measures are crucially important to guarantee the longevity of the project as the codebase and number of contributors keep growing.
The GitHub repository not only serves as a code hosting platform, but also facilitates discussion, interaction and collaboration between all contributors through its issue and milestone tracker, and the project boards.

\subsection*{Data longevity}
With data longevity we refer here to the possibility of accessing data produced with earlier code versions from newer versions.
This is an important aspect of any data infrastructure and it is particularly critical to AiiDA.
Indeed, AiiDA aims at improving the reproducibility of computational science simulations and thus data should be accessible also years after it has been produced.
However, AiiDA also focuses on being performant for high-throughput workloads (see ``Performance''), which often requires changes to existing data layouts.

To guarantee the longevity of data, AiiDA can automatically migrate data across code versions without the need of human intervention.
Particular care has been devoted to implement robust database schema and data migrations for both database backends (Django and SQLAlchemy, see ``Database abstraction and querying language'') with 42 migrations per backend implemented in v1.0.
These migrations ensure that early AiiDA users can seamlessly migrate their databases generated years ago to the most recent version without losing any of their data or provenance.
Additionally, to guarantee that all migrations are correct and thus avoid the potentially irreversible corruption of user data, the integrity of each migration is verified by automated individual unit tests as described previously.
While the development and maintenance effort for these migrations is significant, they are essential for data longevity and with that, for the uptake of the platform by users, by giving them confidence that their data will be accessible in the future.
Similar migrations and the corresponding tests have been implemented for AiiDA archive files that contain exported (parts of) AiiDA databases.
Since these archive files are not concerned with the database schema but merely with the data itself, the database migrations cannot be reused, but instead separate migrations in Python had to be developed.

A final contribution to AiiDA's data longevity is the compatibility of AiiDA 1.0 with both Python 2 and Python 3.
Despite Python 2 reaching its end-of-life in January 2020, AiiDA 1.0 supports this version for another 6 months, thereby providing a grace period for users and developers to upgrade their code and plugins to Python 3.

\subsection*{Interoperability}
Through its plugin system, AiiDA forms a natural interoperability interface for various simulation codes that is centred around the reproducibility provided by the provenance graph.
Other interoperability requirements involve importing and converting data and interfacing to existing libraries.
Indeed, we try not to reimplement features provided by existing robust codes, but rather we provide interfaces to them.
Specifically, AiiDA comes with built-in support for various widespread computational materials science libraries for data analysis and conversion such as ASE~\cite{HjorthLarsen:2017}, pymatgen~\cite{Ong:2013}, spglib~\cite{Atsushi:2018} and seekpath~\cite{Hinuma:2017}.
Additionally, tools are provided to directly interface with existing databases, such as ICSD~\cite{Belsky:2002}, COD~\cite{Grazulis:2012}, TCOD~\cite{Merkys:2017} and OQMD~\cite{Kirklin:2015}, and import the data into an AiiDA database.
Finally, the Materials Cloud dissemination platform~\cite{Talirz:2020} is fully integrated with AiiDA and allows the interactive exploration of any AiiDA database through a graphical user interface.

\subsection*{Outreach}
\label{sec:communitybuilding:outreach}
In addition to the aspects outlined earlier, we actively pursued various activities to strengthen the user and developer community of AiiDA.
Notably, we organised a significant number of events (see \href{http://www.aiida.net/events}{aiida.net/events}), with 14 tutorials, schools and workshops over the 2017-2019 period.
These targeted both new users, introducing the code and the concepts of provenance and reproducible workflows, as well as advanced developers (with yearly coding weeks, and events aiming to provide direct support to AiiDA plugin developers).

To broaden the accessibility of the educational material presented during these events beyond people attending in person, the tutorial texts are made available online~(\href{https://aiida-tutorials.readthedocs.io/en/latest/}{aiida-tutorials.readthedocs.io}) and live recordings of presentations are uploaded to the Materials Cloud~\cite{Talirz:2020} Learn section~(\href{https://www.materialscloud.org/learn/}{materialscloud.org/learn/}).
The tutorial texts are distributed together with virtual machines, based on the Quantum Mobile~\cite{Talirz:2020}, which contain preinstalled and preconfigured versions of AiiDA, its plugins, and the simulation codes and data, such that no setup is required to follow the tutorial.
With this innovative approach, the barrier to access and learn the code is significantly lowered, and interested new users can quickly try out AiiDA and understand if it suits their research needs.
In addition to the interactive tutorials, the code comes with extensive online documentation~(\href{https://aiida-core.readthedocs.io/en/latest/}{aiida-core.readthedocs.io}) and a mailing list is operated by the core developers to provide direct user support~(\href{https://groups.google.com/forum/#!forum/aiidausers}{groups.google.com/forum/\#!forum/aiidausers}).

Thanks to these efforts, AiiDA has seen a rise in its use and adoption for research projects~\cite{Mounet:2018,Kahle:2020,Mercado:2018,Prandini:2018,Vitale:2019}.
The community of plugin developers has also grown substantially and as of March 2020 the plugin registry (see ``The plugin system'') hosts 47 plugin packages, covering over 90 different simulation codes and providing around 80 workflows.
This developer community has also been strengthened by the targeted events that we have organised, for example to help developers migrate their plugins to AiiDA v1.0, to extend support to Python 3, and to design common workflows and interfaces for codes implementing similar methods.

Finally, as AiiDA grows, we have now standardised the approach to provide concrete ideas for improvement and further extensions of AiiDA by the community at large.
Inspired by the concept of the Python Enhancement Proposals (PEPs) (\href{https://www.Python.org/dev/peps/}{Python.org/dev/peps/}), we have implemented a repository to host new AiiDA Enhancement Proposals (AEPs) (\href{https://github.com/aiidateam/AEP}{github.com/aiidateam/AEP}) and a standardised protocol to provide new suggestions.
Discussions on each suggested AEP are facilitated by the GitHub platform and remain available also after approval, allowing to go back and review the reasoning that justified specific design decisions.
AEPs provide a way to extend the discussions on the future roadmap of the code to all interested users, beyond the pool of core developers.

\section*{Conclusions}
\label{sec:conclusions}
We have presented AiiDA 1.0, an open-source high-throughput scalable simulation informatics infrastructure implemented in Python with a strong focus on automated data provenance.
Specifically, we have highlighted the difference in design of the recently released version 1.0 and that of earlier versions, that now makes the solution scalable towards exascale high-throughput computational loads.
AiiDA 1.0 can easily sustain tens of thousands of processes per hour and dispatch them on a broad range of computing resources, from local computers to large high-performance supercomputers.
The key change to achieve this goal has been the redesign of the engine from a polling-based to an event-based paradigm.
The engine can be scaled on demand to an arbitrary number of workers that operate independently, with communication with and among them made possible by the RabbitMQ message broker.

In addition to the engine, significant improvements have been made to the design and implementation of the provenance graph.
The concept of workflows is now fully integrated into the provenance graph, including now also the logical steps in the data provenance.
The implementation of the provenance graph has been optimised and made more efficient by migrating the storage of node properties to native JSONB column types and by computing the transitive closure on the fly instead of storing it in a static table.
The \code{QueryBuilder} is a powerful tool that allows users to inspect their provenance graph and extract information, without having to be able to write SQL.
The tool provides a simple Python syntax that is independent of the ORM backend used for the implementation.
The REST API provides yet another way of extracting data from an AiiDA provenance graph, which is especially useful when direct access to the machine running the instance is not available.
Crucially, despite this myriad of improvements and changes in the code compared to early versions, existing data (and their provenance) can be automatically migrated and therefore are guaranteed to remain compatible with the latest version of AiiDA.

Finally, the plugin system makes AiiDA a flexible tool interoperable with any simulation software.
It is not just limited to calculation plugins; users can create their own data types, command line interface extensions and share their workflows.
The plugin registry allows developers to register their plugin packages and other users to discover them, which directly fosters a lively community of developers, that has grown beyond the original field of application in materials science, supporting now over 90 codes spanning many fields of research such as mechanical engineering, chemistry and physics.
Thanks to advanced tools available within the Python ecosystem, installing the plugin packages can be performed with a single command, automatically registering the included plugins with AiiDA.
Through these tools, AiiDA provides a platform and community that are geared towards making computational science more reproducible.

\section*{Methods}
\label{sec:methods}
\subsection*{\label{app:processrunners}Process runners and the AiiDA daemon}
When a process is launched, an instance of the relevant \code{Process} subclass is created and assigned to a \define{process runner} (or simply runner), which runs it to completion.
Each runner has an internal event loop, which allows it to run multiple AiiDA processes concurrently in a single thread using coroutines (i.e., functions that can yield control during execution when they need to wait for some long-running action to happen; the event loop can then give control to other coroutines that had previously yielded and are ready to continue).
Additionally, each runner has access to a \define{persister}, which implements the logic to serialise and store the state of a process as a checkpoint (in the specific implementation of AiiDA, by serialising the process instance to YAML format and storing it in the database as an attribute of the corresponding node).
This mechanism allows interrupted processes to be restarted, eliminating the need to re-execute code that was already run even if the runner is completely stopped.

The simplest implementation of the runner is the local one, meaning that the process is executed in the current Python interpreter and blocks it until all processes have finished.
However, this approach does not scale well for large numbers of processes.
AiiDA therefore implements daemon runners, that operate exactly like local runners, except that they are spawned in a separate system process that is ran in the background.
We stress here the difference between an AiiDA process, i.e., an entity specific to AiiDA that has data as inputs and as outputs; and a system process, e.g., a Windows or Unix process, that represents the execution of an executable in a computer operating system.
Each daemon runner is fully independent from the others and special care has been devoted in the implementation to allow to run multiple runners in parallel without
concurrency issues when accessing the database.
Given that the daemon runners operate in separate system processes, a message broker is employed to allow communication with and among the runners, as discussed in the next section.

Daemon runners are spawned and controlled by a single main daemonised process, implemented by the \name{circus} library (\href{https://circus.readthedocs.io/en/latest/}{circus.readthedocs.io}).
AiiDA's \code{verdi} command line interface provides easy commands to start and stop the daemon, inspect the status of the daemon runners it controls, and to increase or reduce the number of active runners.

\subsection*{\label{app:rabbitmq}Process communication via the RabbitMQ message broker}
Communication with the runners and among runners is provided by the RabbitMQ message broker (\href{https://www.rabbitmq.com}{rabbitmq.com}).
Each daemon runner subscribes to a special task queue on the message broker to which newly submitted processes are added as \define{tasks}.
The contents of the task queue are persisted to disk by the broker, so that even after an interruption (for example a machine restart), uncompleted tasks can be reloaded and resent to the runners.
The combination of queue persistence, together with the automatic checkpointing of AiiDA processes performed by the engine, ensures that interrupted processes can continue from the last checkpoint, reducing the loss of computational time to the minimum.

The implementation of RabbitMQ and the configuration of the task queue guarantee that each task is only sent to exactly one runner at a time and will eventually be executed.
A task is kept in the queue until it is explicitly acknowledged as completed by the runner that received it.
The broker monitors subscribed runners by sending a periodic ``heartbeat'' call.
If the runner fails to respond within a certain interval twice consecutively, the runner is considered unreachable and tasks assigned to it are redistributed.
To prevent runners from missing the heartbeat call while the main thread is under heavy load, all communication with the broker is performed on a separate thread.
In order to avoid race conditions in database updates, the communication thread has no access to the database but only schedules callbacks on the event loop of the main thread, and forwards broadcasts to the broker.

When a runner receives a task to run a process, it subscribes to a dedicated channel for that specific process.
In such a way, interaction with live processes (e.g., to pause, restart or kill them) is possible by sending remote procedure calls (RPCs) over the appropriate channel.
Conversely, processes can also emit state changes through their runner's communicator.
These are broadcast to subscribed listeners (e.g., a parent workflow waiting for all of its subworkflows to finish) that can then respond with an appropriate action.
This event-based model makes the AiiDA engine able to reply almost instantaneously to events, without the need to periodically poll and check the process states, which would be much more CPU-intensive and less reactive.

\section*{Performance}
\label{sec:performance}
In this section, we examine the performance of the database and the engine.
Where applicable, a comparison is drawn with earlier versions of AiiDA~\cite{Pizzi:2016}, to illustrate how the new design has improved performance.

The database schema of AiiDA 1.0 includes many improvements and optimisations with respect to the schema published in Ref.~\cite{Pizzi:2016}.
Here we highlight two changes that have had the greatest impact on storage and query efficiency.
First, in section ``EAV replaced by JSONB'' we describe how the schema of node attributes has been changed from a custom entity-attribute-value solution to a native JSON binary (JSONB) format.
Second, section ``On-the-fly transitive closure'' details how the transitive-closure, originally a statically generated table, has been replaced with one that is generated on the fly in memory.

\subsection*{EAV replaced by JSONB}
\label{sec:database-performance:jsonb}
As described in ``The provenance model'', the attributes of a node are stored in a relational database.
The exact schema for these attributes depends on the node type and cannot be statically defined, which is in direct conflict with the \emph{modus operandi} of relational databases where schemas are rigorously defined \emph{a priori}.
This limitation was originally overcome by implementing an extended entity-attribute-value (EAV) table that allowed storing arbitrarily nested serialisable attributes in a relational database\cite{Pizzi:2016}.
While a successful solution, it comes with an increased storage cost and significant overhead in the (de-)serialisation of data, reducing the querying efficiency.

As storing semi-structured data is a common requirement for many applications, PostgreSQL added support for a native JSON and JSONB datatype as of v9.2~(\href{https://www.postgresql.org/docs/current/static/release-9-2.html}{postgresql.org/docs/current/static/release-9-2.html}) and v9.4~(\href{https://www.postgresql.org/docs/current/static/release-9-4.html}{postgresql.org/docs/current/static/release-9-4.html}), respectively, which is an efficient storage and indexing format~(\href{https://www.postgresql.org/docs/current/static/datatype-json.html}{postgresql.org/docs/current/static/datatype-json.html}).
In AiiDA 1.0, the custom EAV implementation for node attributes has been replaced with the native JSONB provided by PostgreSQL, which yields significant improvements in both storage cost and query efficiency.

\begin{figure*}[t!]
    \centering
    \includegraphics[width=0.8\linewidth]{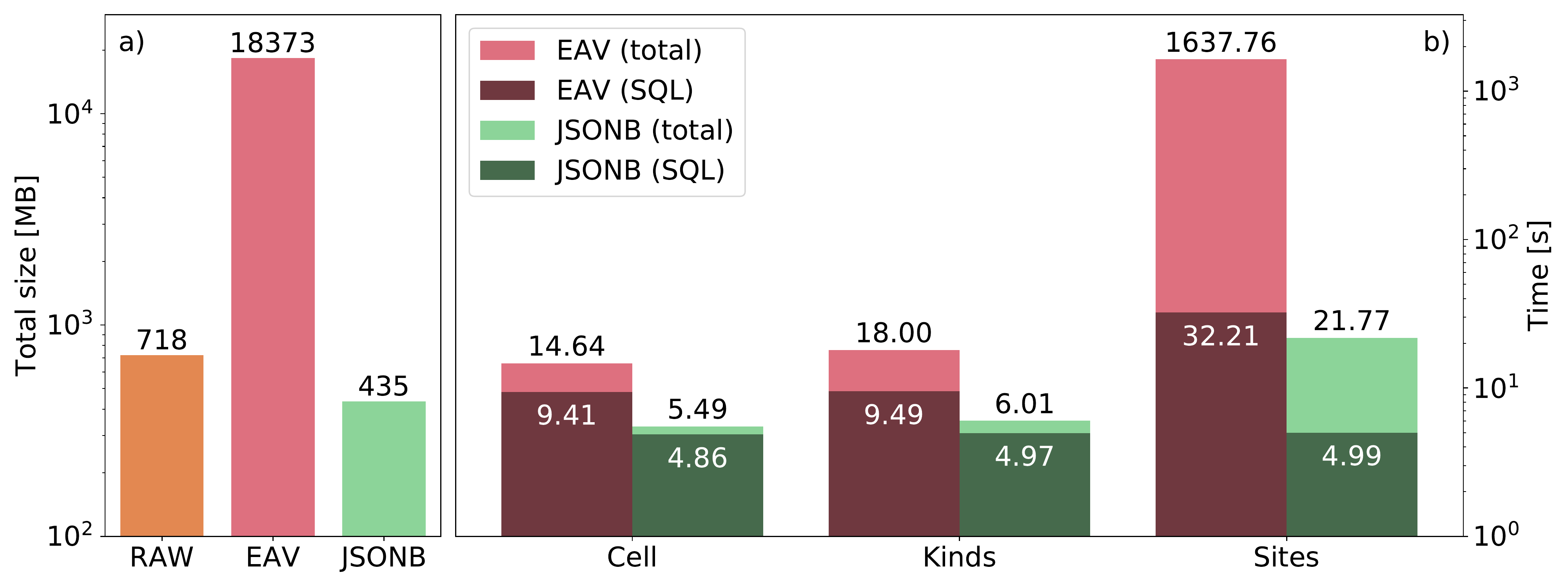}
    \caption{
        Comparison in a log scale of the space requirements and time to solution when querying data with the two AiiDA ORM backends.
        (a) Space needed to store $10\,000$ structure data objects as raw text files, using the existing EAV-based schema and the new JSON-based schema.
        The reduced space requirements of the JSON-based schema with respect to the raw text files are due to, among other things, white-space removal.
        The JSONB schema reduces the required space by a factor of $1.5$ compared to the raw file size and a factor of $25$ compared to the EAV-based schema.
        (b) Time for three different queries that return attributes of different size for the same set of nodes.
        The benchmarks are run on a cold database, meaning that the database caches are emptied before each query.
        We indicate separately the database query time (SQL time) and the total query time which includes also the construction of the Python objects in memory.
        The total query time of the \emph{site} attributes in the JSONB format is $75$ times smaller compared to the equivalent query in the EAV format.
        The SQL time for the same query is roughly $6.5$ times smaller for the JSONB version of SQL query compared to the EAV version of the query.
    }
    \label{fig:bench1}
\end{figure*}

The replacement of EAV by JSONB significantly reduces storage cost in two ways: \textit{(a)} the data itself is stored more compactly as it is reduced from an entire table to a single column and \textit{(b)} database indexes can be removed while still providing a superior query efficiency.
Fig.~\ref{fig:bench1}(a) shows the space occupied when storing $10\,000$ crystal structures, comparing the size of the raw files on disk, and with their content stored in the EAV and JSONB schema.
In the case of raw files, the XSF format~(\href{http://www.xcrysden.org/doc/XSF.html}{xcrysden.org/doc/XSF.html}) was used since it contains only the information that is absolutely necessary.

This benchmark was performed on a PostgreSQL 9.6.8 database using the ORM backends as implemented in AiiDA 1.0.
When comparing the EAV format to the JSONB format, a decrease in storage space of almost two orders of magnitude becomes apparent.
The space gains of the new format do not only apply to the occupied space on disk, but also to the amount of data transferred when querying JSON fields, as shown in Tab.~\ref{tab:query_results_space}.
This effect is, however, only a part of the increase in query efficiency thanks to the JSONB schema.
\begin{table}
    \begin{center}
        \begin{tabular}{c r r r r}
            \toprule
            \multirow{2}{*}{Attribute} & \multicolumn{2}{c}{EAV} & \multicolumn{2}{c}{JSONB} \\ \cmidrule(lr){2-5}
                  & Rows & Size (MB) & Rows & Size (MB) \\
            \cmidrule(lr){1-5}
            Cell  &    255,671  &   71 & 19,667 &   8 \\
            Kinds &    537,303  &  150 & 19,667 &  11 \\
            Sites & 15,352,241  & 4412 & 19,667 & 201 \\
            \bottomrule
        \end{tabular}
        \caption{Result size and number of rows of attribute queries presented in Fig~\ref{fig:bench1}(b) on a database table of $300\,000$ crystal structures.}
        \label{tab:query_results_space}
    \end{center}
\end{table}

Using the JSONB-based format also carries significant speed benefits.
These mainly come from the more compact JSON-based schema with respect to the EAV schema, as described in the previous section.
This results in \textit{(a)} less transferred data from the database to AiiDA, and \textit{(b)} a reduced cost of deserialising the raw query result into Python objects.

Fig.~\ref{fig:bench1}(b) shows benchmarks carried out with PostgreSQL 10.10 on an AiiDA database generated for a research paper~\cite{Mounet:2018} which contains $7\,318\,371$ nodes.
The benchmarks were carried out on a subset of $300\,000$ crystal-structure data nodes on a machine with an Intel i7-5960X CPU with 64GB of RAM.
Three different kind of attributes were queried: \textit{cell}, \textit{kinds} and \textit{sites}.
The \emph{cell} is a $3\times 3$ array of floats, \emph{kinds} contain information on atomic species (therefore, typically there is one kind per chemical element contained in the structure), while there is a \emph{site} per atom, explaining the increase in result sizes as shown at Table~\ref{tab:query_results_space}.

Due to the specific format of the EAV schema, more rows need to be retrieved for every crystal structure data node.
The effect of the different result size is visible both in the SQL time (reflecting the time to perform the query and to get the result from the database) and in the total amount of time spent which includes the deserialisation of raw query results into Python objects.
As shown in Fig~\ref{fig:bench1}(b), the total query time of the \emph{site} attributes in the JSONB format is $75$ times smaller than the equivalent query in the EAV format.
The SQL time for the same query is roughly $6.5$ times smaller for the JSONB version of SQL query compared to the EAV version of the query.
The increased final speedup at the Python level is due to the fact that in the EAV based schema there is the overhead of serialising the attributes at the Python level, which is largely avoided in a JSON-based schema.

\subsection*{On-the-fly transitive closure}
\label{sec:database-performance:tc}
Very often, when querying the provenance graph one is only interested in the neighbours directly adjacent to a certain node.
However, some use cases require to traverse the graph taking multiple hops in the data provenance to find a specific \define{ancestor} or \define{descendant} of a given node.
To make the queries for ancestors and descendants at arbitrary distance as efficient as possible, early versions of AiiDA computed and stored the \define{transitive closure} (TC) of the graph (i.e., the list of all available paths between any pair of nodes) in a separate database table.
Storing these paths in a dedicated database table with appropriate indexes allowed AiiDA to query for ancestors and descendants with time complexity $\mathcal{O}(1)$ independent of the graph topology and the number of hops.

\begin{figure*}
    \centering
    \includegraphics[width=\linewidth]{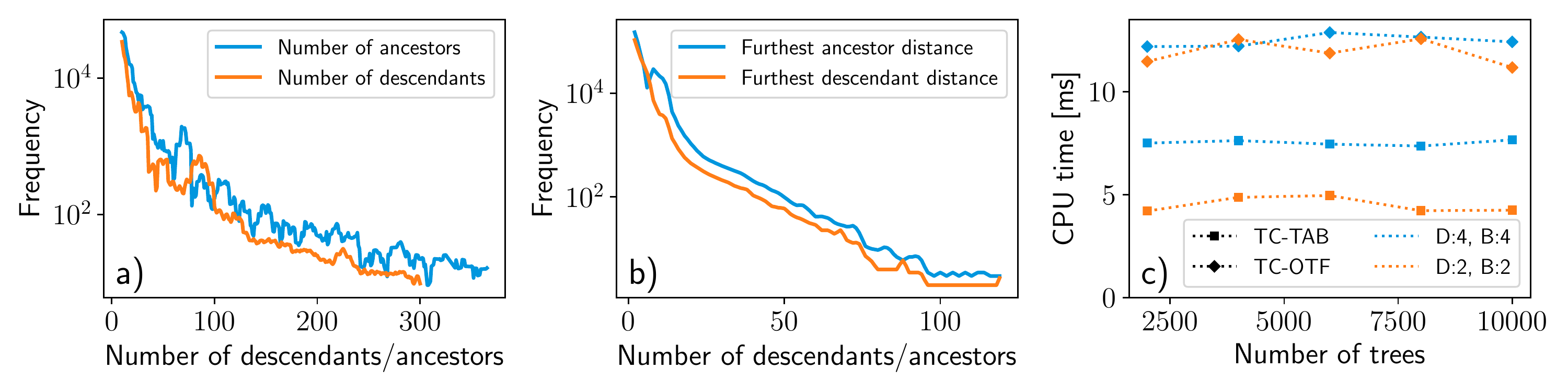}
    \caption{
        Analysis of a sample of one million nodes of the AiiDA graph published in Ref.~\cite{Mounet:2018}.
        (a) Frequencies of the number of ancestors and descendants of all nodes.
        (b) Frequencies of the number of hops, i.e., the distance to reach the farthest ancestor/descendant.
        (c) Required CPU time when querying for all descendants of $50$ top-level nodes in a graph that consists of a number of binary trees of breadth $B$ and depth $D$ using the transitive closure on-the-fly (TC-OTF, diamonds) or the explicitly tabulated transitive closure (TC-TAB, squares).
    }
    \label{fig:aiida_dbpath_statistics}
\end{figure*}

However, the typical size of the TC table is significant even for moderately sized provenance graphs, and quickly has an adverse effect on the general performance of the database.
For example, a subset of just one million nodes from the database generated in Ref.~\cite{Mounet:2018} has 226 million rows in the TC table, corresponding to $200$GB on disk.
In addition to the raw disk storage cost, the time needed to store a new link also increases, as the TC is updated with automatic triggers at each update of the links table.
This becomes more expensive as the table grows because table indexes need to be updated as well.
AiiDA 1.0 replaces the TC explicitly stored in a table with one that is computed lazily, or \define{on-the-fly} (OTF), whenever ancestors or descendants of a node are queried for.
This is implemented in the query builder via SQL common table expressions to recursively traverse the DAG.
The OTF method greatly reduces the time required to store new links and does not require any disk space for storing the TC, albeit at the cost of slightly slower queries.
However, the impact on the efficiency of the recursive queries for AiiDA provenance graphs is minimal, since the typical graph topology is relatively shallow and often composed of (almost) disjoint components.
This can be seen in Fig.~\ref{fig:aiida_dbpath_statistics}(a)-(b) that shows frequency graphs capturing the topology of a subgraph of the database of Ref.~\cite{Mounet:2018} composed of one million nodes.
Indeed, the vast majority of nodes only have a handful of ancestors and descendants and these can be reached in a relatively small number of hops.

To compare the performance of the explicit and lazy implementations of the TC, we performed benchmarks on multiple graphs with topologies comparable to typical AiiDA provenance graphs.
Each graph consists of $N$ binary trees with a depth and breadth (the number of downward branches and the number of outward going edges from each vertex, respectively) of $2$ or $4$.
The benchmark records the total time it takes to query for all descendants of $50$ top-level nodes using either the explicit TC table (TC-TAB) or the on-the-fly TC (TC-OTF).
Fig.~\ref{fig:aiida_dbpath_statistics}(c) clearly shows that in both cases the number of trees does not affect the query efficiency.
Moreover, as the depth and breadth of the graph increases, the TC-TAB query time increases.
In contrast, for the TC-OTF, the topology of the graph has little impact on the query time.
Note that this holds for these particular topologies, which match that of typical AiiDA provenance graphs, but is not necessarily the case for more complex graph topologies.
Finally, as expected, the TC-TAB is faster than the TC-OTF, albeit by just a factor of two.
We deem this increased cost more than acceptable, given the considerable savings in storage space provided by the TC-OTF, the performance independence from the graph topology and the faster storage of new links.
For these reasons, all recent versions of AiiDA implement only the TC-OTF.

\subsection*{Event versus polling-based engine}
\label{sec:engine-performance}
\begin{figure}[tbp]
    \centering
    \includegraphics[width=0.45\textwidth]{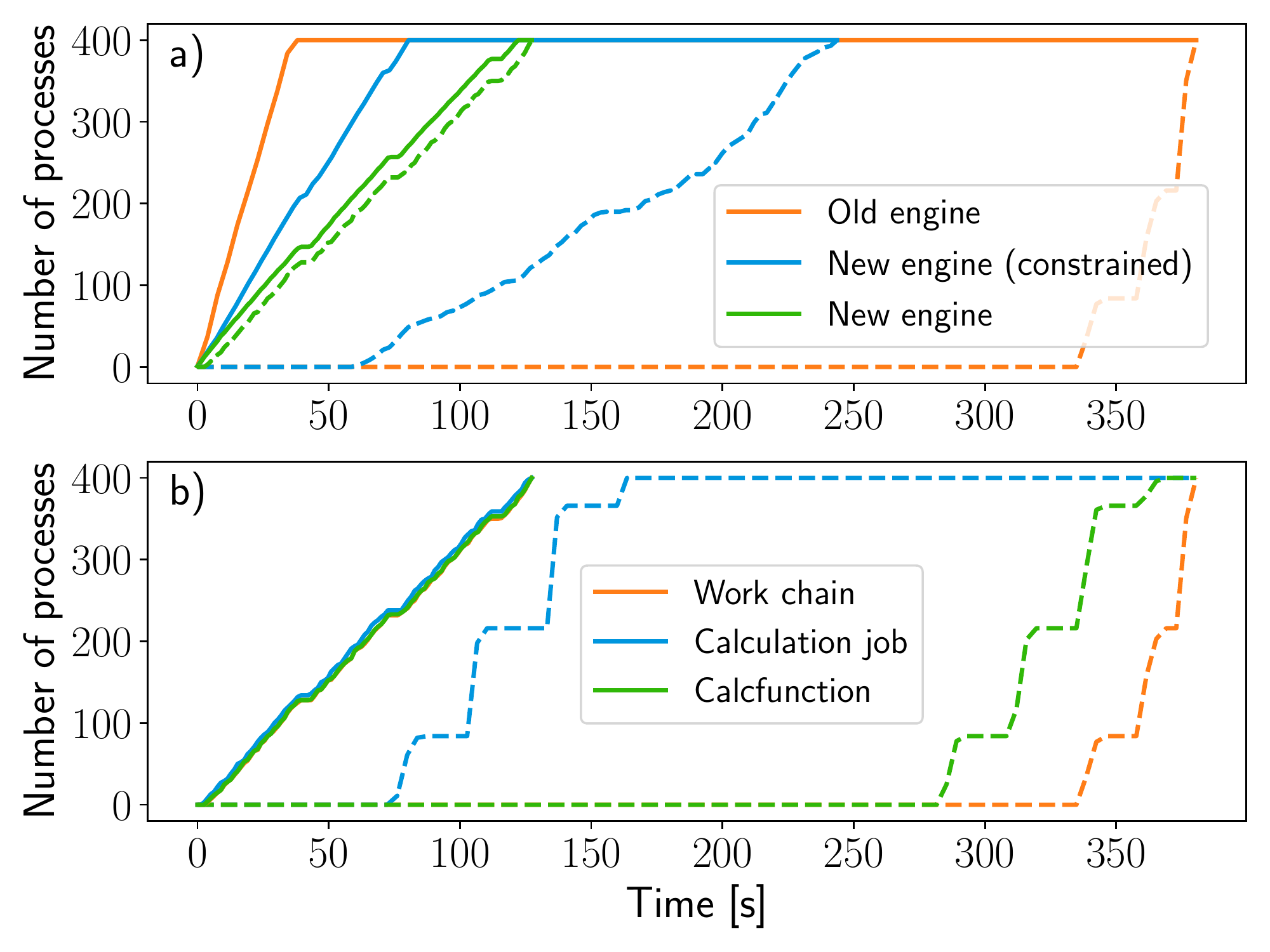}
    \caption{
        Process submission and completion rates for the old and new engine.
        (a) Number of submitted (solid lines) and completed (dashed lines) processes over time for the new engine (both with optimised parameters and with artificial constraints, see text) and the old engine.
        The submission of the old engine is slightly faster, but despite this the completion rate of the new engine is clearly higher, even under constrained conditions.
        (b) Number of completed processes for the old (dashed lines) and new (solid lines) engine, decomposed in the separate (sub)processes.
        The polling-based nature of the old engine is clearly reflected in the stepwise behaviour of the completion rate with processes being finalised in batches.
        In contrast, the curves for the new engine, due to its event-based design, are smooth and closely packed together, indicating processes being executed in a continuous fashion.
    }
    \label{fig:performance-engine}
\end{figure}

To evaluate the performance of the event-based engine of AiiDA v1.0 compared to the polling-based one of earlier versions, we consider an example work chain that performs simple and fast arithmetic operations.
The work chain first computes the sum of two inputs by submitting a \code{CalcJob} and then performs another addition using a \code{calcfunction}.
For the \code{CalcJob}, the \code{ArithmeticAddCalculation} implementation is used, which wraps a simple Bash script that sums two integers.
Each work chain execution then corresponds to the execution of three processes (top work chain, a calculation job and a calculation function) and is representative of typical use cases.
For each benchmark, $400$ work chains are submitted to the daemon and the rate of submission and process completion is recorded, as shown in Fig.~\ref{fig:performance-engine}.
These benchmarks were performed on a machine with an Intel Xeon E5-2623 v3 CPU with 64GB of RAM.
Beside the results obtained for the old and new engine using optimised parameters (number of workers, transport intervals, \ldots), for a fair comparison and to highlight the effect of different engine types, in Fig.~\ref{fig:performance-engine}(a) we also show the results for the new engine with some artificial constraints.
In particular, we run the new engine with four workers only (which is roughly comparable with the old engine, with four independent tasks for submitting, checking queued jobs, retrieving files, and processing work chain steps) rather than twelve.
Additionally we set a minimal interval between connections of 5 seconds in the new daemon to simulate the polling behaviour (with default polling time of 5 seconds) of the old daemon, despite all calculation jobs being run on the local host where an interval of zero is optimal.

Fig.~\ref{fig:performance-engine}(a) shows that the submission rate of the old engine is slightly faster compared to the new engine, because the procedure was significantly simpler with no communication with remote daemon workers.
Nevertheless, the total completion time of the new engine (even in the constrained configuration) is shorter, with the optimised new engine completing all processes three times faster than the old one in this simple example.
Additionally, for the old engine all work chains complete towards the end of the time window at roughly the same time, in a few discontinuous jumps because of the polling-based architecture.
In contrast, the completion rate of the event-based engine is much smoother (beside being faster in general) because of the continuous operating character of the new engine, with processes managed concurrently and executed immediately, without waiting for the next polling interval.

The concurrency of the new daemon is highlighted even more in Fig.~\ref{fig:performance-engine}(b) where we compare the completion time for the new (optimised) daemon and old one, showing independently the completion for the top work chain and each of the two subprocesses.
The reason of the large delay in the old engine is because, even though the workflow only runs two subprocesses, the internal logic consists of multiple steps, where only one is processed per polling interval.
In contrast, the new engine executes all workflow steps in quick succession without interruption.

We stress that the efficiency improvements of the new engine are even larger in real high-throughput situations, since the daemon is never idle between polling intervals.
Most importantly, the new daemon is scalable and the number of daemon workers can be increased dynamically to distribute heavy work loads.
This effect is made visible in Fig.~\ref{fig:performance-engine}(a), where the optimised new engine (with 12 workers and without connection delay) completes all processes in half the time required by the constrained one.
The effective throughput of the new engine for this experiment, which was run on a modest work station, amounts to roughly $35\,000$ processes per hour.
Due to the scalable design of the new engine, this rate can be easily increased by running more daemon runners on a more powerful machine.

\subsection*{Caching}
The storing of complete data provenance as described in ``The provenance model'' does not only guarantee the reproducibility of results, it can also reduce the unnecessary repetition of calculations.
If the engine is asked to launch a calculation, it can first check in the database if a calculation with the exact same inputs has already been performed.
In that case, the engine can simply reuse the outputs of the completed calculation saving computational resources.
This mechanism is referred to as \define{caching} in AiiDA and users can activate it for all calculation types or only for specific ones.

To rapidly find identical calculations in the database, one needs an efficient method to determine whether two nodes are equivalent.
For this purpose, AiiDA uses the concept of hashing, where the whole content of a node is mapped onto a single short hexadecimal string.
In AiiDA we employ the cryptographic BLAKE2b algorithm (\href{https://blake2.net}{blake2.net}), which has a relatively low computational cost combined with an overwhelmingly unlikely probability of hash collisions.
The latter property means that any two nodes with the same hash can be assumed to be identical.
The content of a node that is included in the computation of its hash consists of the immutable node attributes and the file repository contents.
In addition, for a calculation node the hashes of all its inputs are also included, such that looking for calculations with identical inputs can be done merely by looking at the hash of the calculation itself.

As soon as a node is stored and it becomes immutable, its hash is computed and stored as a node property, making it queryable.
When the engine is asked to launch a new calculation, it first computes its hash and searches the database for an existing node with the same hash.
If an identical calculation is found and caching is enabled, the engine simply clones the output nodes of the existing calculation and links them to the new calculation.
This saves valuable computational resources and results in the same provenance graph as if the calculation had actually been run.
Nevertheless, specific extra properties are added to indicate which calculation was used as the cache source, making it possible to identify cached calculations (mostly for debugging purposes).

The concept of caching is especially powerful when developing and running complex workflows consisting of many calculations.
If any calculation fails, the workflow that launched it fails as well.
If the error can be resolved (e.g., because it was due to a bug in the workflow), the workflow can be fixed and simply rerun from scratch:
thanks to the caching mechanism, it will effectively continue from where it previously failed without repeating successful calculations.

\section*{Code availability}
\label{sec:codeavailability}
The source code of AiiDA is released under the MIT open-source license and is made available on GitHub (\href{https://github.com/aiidateam/aiida-core}{github.com/aiidateam/aiida-core}).
It is also distributed as an installable package through the Python Package Index (\href{https://pypi.org/project/aiida-core}{pypi.org/project/aiida-core}).

\section*{Data availability}
\label{sec:dataavailability}
The data used to create Fig~\ref{fig:bench1}, Fig.~\ref{fig:aiida_dbpath_statistics} and Tab.~\ref{tab:query_results_space} for the analysis of the database performance come from Ref.~\cite{Mounet:2018}.
The raw data is available on the Materials Cloud Archive~\cite{web:Mounet:MCA:2018}.

The data used to create Fig.~\ref{fig:performance-engine}, along with the scripts that were used for the creation and analysis of the data, are available on the Materials Cloud Archive~\cite{web:Huber:MCA:2020}.

\section*{Acknowledgments}
This work is supported by the MARVEL National Centre for Competency in Research funded by the Swiss National Science Foundation (grant agreement ID~51NF40-182892), the European Centre of Excellence MaX ``Materials design at the Exascale'' (grant no.~824143) and by the Swiss Platform for Advanced Scientific Computing (PASC).

Additional support was provided by the ``MaGic'' project of the European Research Council (grant agreement ID~666983), the swissuniversities P-5 ``Materials Cloud'' project (grant agreement ID~182-008), the ``MARKETPLACE'' H2020 project (grant agreement ID~760173), the ``INTERSECT'' H2020 project  (grant agreement ID~814487), the ``NFFA'' H2020 project (grant agreement ID~654360) and the ``EMMC'' H2020 project (grant agreement ID~723867).

We would like to thank the following people for code contributions, bug fixes, improvements of the documentation, and useful discussions and suggestions: Oscar D. Arbel\'aez-Echeverri, Michael Atambo, Valentin Bersier, Marco Borelli, Jocelyn Boullier, Jens Br\"oder, Ivano E. Castelli, Keija Cui, Vladimir Dikan, Marco Dorigo, Y.-W. Fang, Espen Flage-Larsen, Marco Gibertini, Daniel Hollas, Eric Hontz, Jianxing Huang, Christoph Koch, Ian Lee, Daniel Marchand, Antimo Marrazzo, Simon Pintarelli, Gianluca Prandini, Philipp R\"u{\ss}mann, Philippe Schwaller, Ole Sch\"utt, Christopher Sewell, Andreas Stamminger, Atsushi Togo, Daniele Tomerini, Nicola Varini, Jason Yu, Austin Zadoks, Bonan Zhu and Mario Zic.

\section*{Author contributions}
The initial contributions to AiiDA are listed in Ref.~\cite{Pizzi:2016}; here we list the developments and contributions since then.
GP and NMa supervised and coordinated the project.
MU, GP and SPH designed and MU and SPH implemented the new engine, with contributions from DG in parts of the user interface.
BK designed the use of process functions to track the provenance of Python functions, and GP, NMo, AM, MU and SPH implemented it.
RH, MU and SPH designed and implemented the new daemon.
BK formulated the extension of the provenance graph definition to include link types.
SPH and MU designed the current provenance graph formalising the two provenance levels, incorporating the extension of BK, and implemented it with contributions in both design and implementation by LT, GP, AVY, SZ.
BK developed an initial scheme for JSONB node attributes and their queries via the SQLAlchemy ORM.
SZ oversaw and implemented the abstraction of the backend ORM and the addition of SQLAlchemy support, with contributions by AC, LK and FG.
SZ and SPH implemented the use of JSONB fields also with the Django backend.
MU designed the front-end/back-end ORM separation and SPH, MU, GP, LT, SK, SZ and CJ implemented it.
LK, NM and GP implemented the on-the-fly transitive closure.
LK designed and implemented the QueryBuilder.
LK designed the graph traversal tool and implemented it with FFR.
TM and RH designed and implemented the plugin system.
RH redesigned the command line tool and implemented it with SPH, SZ, AVY and LT.
TM added Python 3 support.
BK suggested the idea of caching of calculations.
DG designed and implemented the caching mechanism.
SZ, GP, CWA and AVY improved the export and import functionality.
FG, SK and EP implemented the REST API.
AM designed and implemented the tools to import crystal structures from external databases and to export them to external databases.
CSA and LT have setup and contributed to the AiiDA Enhancement Proposal (AEP) system.
SPH, SZ, GP, MU, LT, LK, RH, NMo, AVY, CWA, FFR, CSA, CJ, AM, AC, FG, SK and EP are or have been part of the AiiDA core developers team that maintains the software.
NMo, AM and AC integrated various external libraries for data transformation and implemented many command line utilities to export and visualise data.
NMo, GP, FG, SZ, SPH, MU, LT, LK, AVY, RH, CWA, AM, AC, SK, EP contributed to the various tutorials that have been organised.
All authors have read and approved the final version of the manuscript.

\section*{Competing interests}
The authors declare no competing interests.

\section*{Bibliography}
\bibliographystyle{apsrev4-1}
\bibliography{manuscript}

\begin{thebibliography}{31}%
\makeatletter
\providecommand \@ifxundefined [1]{%
 \@ifx{#1\undefined}
}%
\providecommand \@ifnum [1]{%
 \ifnum #1\expandafter \@firstoftwo
 \else \expandafter \@secondoftwo
 \fi
}%
\providecommand \@ifx [1]{%
 \ifx #1\expandafter \@firstoftwo
 \else \expandafter \@secondoftwo
 \fi
}%
\providecommand \natexlab [1]{#1}%
\providecommand \enquote  [1]{``#1''}%
\providecommand \bibnamefont  [1]{#1}%
\providecommand \bibfnamefont [1]{#1}%
\providecommand \citenamefont [1]{#1}%
\providecommand \href@noop [0]{\@secondoftwo}%
\providecommand \href [0]{\begingroup \@sanitize@url \@href}%
\providecommand \@href[1]{\@@startlink{#1}\@@href}%
\providecommand \@@href[1]{\endgroup#1\@@endlink}%
\providecommand \@sanitize@url [0]{\catcode `\\12\catcode `\$12\catcode
  `\&12\catcode `\#12\catcode `\^12\catcode `\_12\catcode `\%12\relax}%
\providecommand \@@startlink[1]{}%
\providecommand \@@endlink[0]{}%
\providecommand \url  [0]{\begingroup\@sanitize@url \@url }%
\providecommand \@url [1]{\endgroup\@href {#1}{\urlprefix }}%
\providecommand \urlprefix  [0]{URL }%
\providecommand \Eprint [0]{\href }%
\providecommand \doibase [0]{http://dx.doi.org/}%
\providecommand \selectlanguage [0]{\@gobble}%
\providecommand \bibinfo  [0]{\@secondoftwo}%
\providecommand \bibfield  [0]{\@secondoftwo}%
\providecommand \translation [1]{[#1]}%
\providecommand \BibitemOpen [0]{}%
\providecommand \bibitemStop [0]{}%
\providecommand \bibitemNoStop [0]{.\EOS\space}%
\providecommand \EOS [0]{\spacefactor3000\relax}%
\providecommand \BibitemShut  [1]{\csname bibitem#1\endcsname}%
\let\auto@bib@innerbib\@empty
\bibitem [{\citenamefont {Ioannidis}\ \emph {et~al.}(2009)\citenamefont
  {Ioannidis}, \citenamefont {Allison}, \citenamefont {Ball}, \citenamefont
  {Coulibaly}, \citenamefont {Cui}, \citenamefont {Culhane}, \citenamefont
  {Falchi}, \citenamefont {Furlanello}, \citenamefont {Game}, \citenamefont
  {Jurman}, \citenamefont {Mangion}, \citenamefont {Mehta}, \citenamefont
  {Nitzberg}, \citenamefont {Page}, \citenamefont {Petretto},\ and\
  \citenamefont {van Noort}}]{Ioannidis:2009}%
  \BibitemOpen
  \bibfield  {author} {\bibinfo {author} {\bibfnamefont {J.~P.~A.}\
  \bibnamefont {Ioannidis}}, \bibinfo {author} {\bibfnamefont {D.~B.}\
  \bibnamefont {Allison}}, \bibinfo {author} {\bibfnamefont {C.~A.}\
  \bibnamefont {Ball}}, \bibinfo {author} {\bibfnamefont {I.}~\bibnamefont
  {Coulibaly}}, \bibinfo {author} {\bibfnamefont {X.}~\bibnamefont {Cui}},
  \bibinfo {author} {\bibfnamefont {A.~C.}\ \bibnamefont {Culhane}}, \bibinfo
  {author} {\bibfnamefont {M.}~\bibnamefont {Falchi}}, \bibinfo {author}
  {\bibfnamefont {C.}~\bibnamefont {Furlanello}}, \bibinfo {author}
  {\bibfnamefont {L.}~\bibnamefont {Game}}, \bibinfo {author} {\bibfnamefont
  {G.}~\bibnamefont {Jurman}}, \bibinfo {author} {\bibfnamefont
  {J.}~\bibnamefont {Mangion}}, \bibinfo {author} {\bibfnamefont
  {T.}~\bibnamefont {Mehta}}, \bibinfo {author} {\bibfnamefont
  {M.}~\bibnamefont {Nitzberg}}, \bibinfo {author} {\bibfnamefont {G.~P.}\
  \bibnamefont {Page}}, \bibinfo {author} {\bibfnamefont {E.}~\bibnamefont
  {Petretto}}, \ and\ \bibinfo {author} {\bibfnamefont {V.}~\bibnamefont {van
  Noort}},\ }\href {\doibase 10.1038/ng.295} {\bibfield  {journal} {\bibinfo
  {journal} {Nature Genetics}\ }\textbf {\bibinfo {volume} {41}},\ \bibinfo
  {pages} {149} (\bibinfo {year} {2009})}\BibitemShut {NoStop}%
\bibitem [{\citenamefont {Peng}(2011)}]{Peng:2011}%
  \BibitemOpen
  \bibfield  {author} {\bibinfo {author} {\bibfnamefont {R.~D.}\ \bibnamefont
  {Peng}},\ }\href {\doibase 10.1126/science.1213847} {\bibfield  {journal}
  {\bibinfo  {journal} {Science}\ }\textbf {\bibinfo {volume} {334}},\ \bibinfo
  {pages} {1226} (\bibinfo {year} {2011})}\BibitemShut {NoStop}%
\bibitem [{\citenamefont {Stoddart}(2016)}]{Stoddart:2016}%
  \BibitemOpen
  \bibfield  {author} {\bibinfo {author} {\bibfnamefont {C.}~\bibnamefont
  {Stoddart}},\ }\href {\doibase 10.1038/d41586-019-00067-3} {\bibfield
  {journal} {\bibinfo  {journal} {Nature}\ } (\bibinfo {year} {2016}),\
  10.1038/d41586-019-00067-3}\BibitemShut {NoStop}%
\bibitem [{\citenamefont {Allison}\ \emph {et~al.}(2016)\citenamefont
  {Allison}, \citenamefont {Brown}, \citenamefont {George},\ and\ \citenamefont
  {Kaiser}}]{Allison:2016}%
  \BibitemOpen
  \bibfield  {author} {\bibinfo {author} {\bibfnamefont {D.~B.}\ \bibnamefont
  {Allison}}, \bibinfo {author} {\bibfnamefont {A.~W.}\ \bibnamefont {Brown}},
  \bibinfo {author} {\bibfnamefont {B.~J.}\ \bibnamefont {George}}, \ and\
  \bibinfo {author} {\bibfnamefont {K.~A.}\ \bibnamefont {Kaiser}},\ }\href
  {\doibase 10.1038/530027a} {\bibfield  {journal} {\bibinfo  {journal}
  {Nature}\ }\textbf {\bibinfo {volume} {530}},\ \bibinfo {pages} {27}
  (\bibinfo {year} {2016})}\BibitemShut {NoStop}%
\bibitem [{\citenamefont {Wilkinson}\ \emph {et~al.}(2016)\citenamefont
  {Wilkinson}, \citenamefont {Dumontier}, \citenamefont {Aalbersberg},
  \citenamefont {Appleton}, \citenamefont {Axton}, \citenamefont {Baak},
  \citenamefont {Blomberg}, \citenamefont {Boiten}, \citenamefont
  {da~Silva~Santos}, \citenamefont {Bourne}, \citenamefont {Bouwman},
  \citenamefont {Brookes}, \citenamefont {Clark}, \citenamefont {Crosas},
  \citenamefont {Dillo}, \citenamefont {Dumon}, \citenamefont {Edmunds},
  \citenamefont {Evelo}, \citenamefont {Finkers}, \citenamefont
  {Gonzalez-Beltran}, \citenamefont {Gray}, \citenamefont {Groth},
  \citenamefont {Goble}, \citenamefont {Grethe}, \citenamefont {Heringa},
  \citenamefont {'t~Hoen}, \citenamefont {Hooft}, \citenamefont {Kuhn},
  \citenamefont {Kok}, \citenamefont {Kok}, \citenamefont {Lusher},
  \citenamefont {Martone}, \citenamefont {Mons}, \citenamefont {Packer},
  \citenamefont {Persson}, \citenamefont {Rocca-Serra}, \citenamefont {Roos},
  \citenamefont {van Schaik}, \citenamefont {Sansone}, \citenamefont
  {Schultes}, \citenamefont {Sengstag}, \citenamefont {Slater}, \citenamefont
  {Strawn}, \citenamefont {Swertz}, \citenamefont {Thompson}, \citenamefont
  {van~der Lei}, \citenamefont {van Mulligen}, \citenamefont {Velterop},
  \citenamefont {Waagmeester}, \citenamefont {Wittenburg}, \citenamefont
  {Wolstencroft}, \citenamefont {Zhao},\ and\ \citenamefont
  {Mons}}]{Wilkinson:2016}%
  \BibitemOpen
  \bibfield  {author} {\bibinfo {author} {\bibfnamefont {M.~D.}\ \bibnamefont
  {Wilkinson}}, \bibinfo {author} {\bibfnamefont {M.}~\bibnamefont
  {Dumontier}}, \bibinfo {author} {\bibfnamefont {I.~J.}\ \bibnamefont
  {Aalbersberg}}, \bibinfo {author} {\bibfnamefont {G.}~\bibnamefont
  {Appleton}}, \bibinfo {author} {\bibfnamefont {M.}~\bibnamefont {Axton}},
  \bibinfo {author} {\bibfnamefont {A.}~\bibnamefont {Baak}}, \bibinfo {author}
  {\bibfnamefont {N.}~\bibnamefont {Blomberg}}, \bibinfo {author}
  {\bibfnamefont {J.-W.}\ \bibnamefont {Boiten}}, \bibinfo {author}
  {\bibfnamefont {L.~B.}\ \bibnamefont {da~Silva~Santos}}, \bibinfo {author}
  {\bibfnamefont {P.~E.}\ \bibnamefont {Bourne}}, \bibinfo {author}
  {\bibfnamefont {J.}~\bibnamefont {Bouwman}}, \bibinfo {author} {\bibfnamefont
  {A.~J.}\ \bibnamefont {Brookes}}, \bibinfo {author} {\bibfnamefont
  {T.}~\bibnamefont {Clark}}, \bibinfo {author} {\bibfnamefont
  {M.}~\bibnamefont {Crosas}}, \bibinfo {author} {\bibfnamefont
  {I.}~\bibnamefont {Dillo}}, \bibinfo {author} {\bibfnamefont
  {O.}~\bibnamefont {Dumon}}, \bibinfo {author} {\bibfnamefont
  {S.}~\bibnamefont {Edmunds}}, \bibinfo {author} {\bibfnamefont {C.~T.}\
  \bibnamefont {Evelo}}, \bibinfo {author} {\bibfnamefont {R.}~\bibnamefont
  {Finkers}}, \bibinfo {author} {\bibfnamefont {A.}~\bibnamefont
  {Gonzalez-Beltran}}, \bibinfo {author} {\bibfnamefont {A.~J.}\ \bibnamefont
  {Gray}}, \bibinfo {author} {\bibfnamefont {P.}~\bibnamefont {Groth}},
  \bibinfo {author} {\bibfnamefont {C.}~\bibnamefont {Goble}}, \bibinfo
  {author} {\bibfnamefont {J.~S.}\ \bibnamefont {Grethe}}, \bibinfo {author}
  {\bibfnamefont {J.}~\bibnamefont {Heringa}}, \bibinfo {author} {\bibfnamefont
  {P.~A.}\ \bibnamefont {'t~Hoen}}, \bibinfo {author} {\bibfnamefont
  {R.}~\bibnamefont {Hooft}}, \bibinfo {author} {\bibfnamefont
  {T.}~\bibnamefont {Kuhn}}, \bibinfo {author} {\bibfnamefont {R.}~\bibnamefont
  {Kok}}, \bibinfo {author} {\bibfnamefont {J.}~\bibnamefont {Kok}}, \bibinfo
  {author} {\bibfnamefont {S.~J.}\ \bibnamefont {Lusher}}, \bibinfo {author}
  {\bibfnamefont {M.~E.}\ \bibnamefont {Martone}}, \bibinfo {author}
  {\bibfnamefont {A.}~\bibnamefont {Mons}}, \bibinfo {author} {\bibfnamefont
  {A.~L.}\ \bibnamefont {Packer}}, \bibinfo {author} {\bibfnamefont
  {B.}~\bibnamefont {Persson}}, \bibinfo {author} {\bibfnamefont
  {P.}~\bibnamefont {Rocca-Serra}}, \bibinfo {author} {\bibfnamefont
  {M.}~\bibnamefont {Roos}}, \bibinfo {author} {\bibfnamefont {R.}~\bibnamefont
  {van Schaik}}, \bibinfo {author} {\bibfnamefont {S.-A.}\ \bibnamefont
  {Sansone}}, \bibinfo {author} {\bibfnamefont {E.}~\bibnamefont {Schultes}},
  \bibinfo {author} {\bibfnamefont {T.}~\bibnamefont {Sengstag}}, \bibinfo
  {author} {\bibfnamefont {T.}~\bibnamefont {Slater}}, \bibinfo {author}
  {\bibfnamefont {G.}~\bibnamefont {Strawn}}, \bibinfo {author} {\bibfnamefont
  {M.~A.}\ \bibnamefont {Swertz}}, \bibinfo {author} {\bibfnamefont
  {M.}~\bibnamefont {Thompson}}, \bibinfo {author} {\bibfnamefont
  {J.}~\bibnamefont {van~der Lei}}, \bibinfo {author} {\bibfnamefont
  {E.}~\bibnamefont {van Mulligen}}, \bibinfo {author} {\bibfnamefont
  {J.}~\bibnamefont {Velterop}}, \bibinfo {author} {\bibfnamefont
  {A.}~\bibnamefont {Waagmeester}}, \bibinfo {author} {\bibfnamefont
  {P.}~\bibnamefont {Wittenburg}}, \bibinfo {author} {\bibfnamefont
  {K.}~\bibnamefont {Wolstencroft}}, \bibinfo {author} {\bibfnamefont
  {J.}~\bibnamefont {Zhao}}, \ and\ \bibinfo {author} {\bibfnamefont
  {B.}~\bibnamefont {Mons}},\ }\href {\doibase 10.1038/sdata.2016.18}
  {\bibfield  {journal} {\bibinfo  {journal} {Scientific Data}\ }\textbf
  {\bibinfo {volume} {3}} (\bibinfo {year} {2016}),\
  10.1038/sdata.2016.18}\BibitemShut {NoStop}%
\bibitem [{\citenamefont {Pizzi}\ \emph {et~al.}(2016)\citenamefont {Pizzi},
  \citenamefont {Cepellotti}, \citenamefont {Sabatini}, \citenamefont
  {Marzari},\ and\ \citenamefont {Kozinsky}}]{Pizzi:2016}%
  \BibitemOpen
  \bibfield  {author} {\bibinfo {author} {\bibfnamefont {G.}~\bibnamefont
  {Pizzi}}, \bibinfo {author} {\bibfnamefont {A.}~\bibnamefont {Cepellotti}},
  \bibinfo {author} {\bibfnamefont {R.}~\bibnamefont {Sabatini}}, \bibinfo
  {author} {\bibfnamefont {N.}~\bibnamefont {Marzari}}, \ and\ \bibinfo
  {author} {\bibfnamefont {B.}~\bibnamefont {Kozinsky}},\ }\href {\doibase
  10.1016/j.commatsci.2015.09.013} {\bibfield  {journal} {\bibinfo  {journal}
  {Computational Materials Science}\ }\textbf {\bibinfo {volume} {111}},\
  \bibinfo {pages} {218} (\bibinfo {year} {2016})}\BibitemShut {NoStop}%
\bibitem [{\citenamefont {Uhrin}\ \emph {et~al.}(2020)\citenamefont {Uhrin},
  \citenamefont {Huber}, \citenamefont {Marzari},\ and\ \citenamefont
  {Pizzi}}]{Uhrin:2020}%
  \BibitemOpen
  \bibfield  {author} {\bibinfo {author} {\bibfnamefont {M.}~\bibnamefont
  {Uhrin}}, \bibinfo {author} {\bibfnamefont {S.~P.}\ \bibnamefont {Huber}},
  \bibinfo {author} {\bibfnamefont {N.}~\bibnamefont {Marzari}}, \ and\
  \bibinfo {author} {\bibfnamefont {G.}~\bibnamefont {Pizzi}},\ }\href@noop {}
  {\bibfield  {journal} {\bibinfo  {journal} {in preparation}\ } (\bibinfo
  {year} {2020})}\BibitemShut {NoStop}%
\bibitem [{\citenamefont {Maffioletti}\ and\ \citenamefont
  {Murri}(2012)}]{Maffioletti:2012}%
  \BibitemOpen
  \bibfield  {author} {\bibinfo {author} {\bibfnamefont {S.}~\bibnamefont
  {Maffioletti}}\ and\ \bibinfo {author} {\bibfnamefont {R.}~\bibnamefont
  {Murri}},\ }in\ \href {\doibase 10.22323/1.162.0143} {\emph {\bibinfo
  {booktitle} {Proceedings of {EGI} Community Forum 2012 / {EMI} Second
  Technical Conference {\textemdash} {PoS}({EGICF}12-{EMITC}2)}}}\ (\bibinfo
  {publisher} {Sissa Medialab},\ \bibinfo {year} {2012})\BibitemShut {NoStop}%
\bibitem [{\citenamefont {Mathew}\ \emph {et~al.}(2017)\citenamefont {Mathew},
  \citenamefont {Montoya}, \citenamefont {Faghaninia}, \citenamefont
  {Dwarakanath}, \citenamefont {Aykol}, \citenamefont {Tang}, \citenamefont
  {heng Chu}, \citenamefont {Smidt}, \citenamefont {Bocklund}, \citenamefont
  {Horton}, \citenamefont {Dagdelen}, \citenamefont {Wood}, \citenamefont
  {Liu}, \citenamefont {Neaton}, \citenamefont {Ong}, \citenamefont {Persson},\
  and\ \citenamefont {Jain}}]{Mathew:2017}%
  \BibitemOpen
  \bibfield  {author} {\bibinfo {author} {\bibfnamefont {K.}~\bibnamefont
  {Mathew}}, \bibinfo {author} {\bibfnamefont {J.~H.}\ \bibnamefont {Montoya}},
  \bibinfo {author} {\bibfnamefont {A.}~\bibnamefont {Faghaninia}}, \bibinfo
  {author} {\bibfnamefont {S.}~\bibnamefont {Dwarakanath}}, \bibinfo {author}
  {\bibfnamefont {M.}~\bibnamefont {Aykol}}, \bibinfo {author} {\bibfnamefont
  {H.}~\bibnamefont {Tang}}, \bibinfo {author} {\bibfnamefont {I.}~\bibnamefont
  {heng Chu}}, \bibinfo {author} {\bibfnamefont {T.}~\bibnamefont {Smidt}},
  \bibinfo {author} {\bibfnamefont {B.}~\bibnamefont {Bocklund}}, \bibinfo
  {author} {\bibfnamefont {M.}~\bibnamefont {Horton}}, \bibinfo {author}
  {\bibfnamefont {J.}~\bibnamefont {Dagdelen}}, \bibinfo {author}
  {\bibfnamefont {B.}~\bibnamefont {Wood}}, \bibinfo {author} {\bibfnamefont
  {Z.-K.}\ \bibnamefont {Liu}}, \bibinfo {author} {\bibfnamefont
  {J.}~\bibnamefont {Neaton}}, \bibinfo {author} {\bibfnamefont {S.~P.}\
  \bibnamefont {Ong}}, \bibinfo {author} {\bibfnamefont {K.}~\bibnamefont
  {Persson}}, \ and\ \bibinfo {author} {\bibfnamefont {A.}~\bibnamefont
  {Jain}},\ }\href {\doibase 10.1016/j.commatsci.2017.07.030} {\bibfield
  {journal} {\bibinfo  {journal} {Computational Materials Science}\ }\textbf
  {\bibinfo {volume} {139}},\ \bibinfo {pages} {140} (\bibinfo {year}
  {2017})}\BibitemShut {NoStop}%
\bibitem [{\citenamefont {Adorf}\ \emph {et~al.}(2018)\citenamefont {Adorf},
  \citenamefont {Dodd}, \citenamefont {Ramasubramani},\ and\ \citenamefont
  {Glotzer}}]{Adorf:2018}%
  \BibitemOpen
  \bibfield  {author} {\bibinfo {author} {\bibfnamefont {C.~S.}\ \bibnamefont
  {Adorf}}, \bibinfo {author} {\bibfnamefont {P.~M.}\ \bibnamefont {Dodd}},
  \bibinfo {author} {\bibfnamefont {V.}~\bibnamefont {Ramasubramani}}, \ and\
  \bibinfo {author} {\bibfnamefont {S.~C.}\ \bibnamefont {Glotzer}},\ }\href
  {\doibase 10.1016/j.commatsci.2018.01.035} {\bibfield  {journal} {\bibinfo
  {journal} {Computational Materials Science}\ }\textbf {\bibinfo {volume}
  {146}},\ \bibinfo {pages} {220} (\bibinfo {year} {2018})}\BibitemShut
  {NoStop}%
\bibitem [{\citenamefont {Babuji}\ \emph {et~al.}(2019)\citenamefont {Babuji},
  \citenamefont {Foster}, \citenamefont {Wilde}, \citenamefont {Chard},
  \citenamefont {Woodard}, \citenamefont {Li}, \citenamefont {Katz},
  \citenamefont {Clifford}, \citenamefont {Kumar}, \citenamefont {Lacinski},
  \citenamefont {Chard},\ and\ \citenamefont {Wozniak}}]{Babuji:2019}%
  \BibitemOpen
  \bibfield  {author} {\bibinfo {author} {\bibfnamefont {Y.}~\bibnamefont
  {Babuji}}, \bibinfo {author} {\bibfnamefont {I.}~\bibnamefont {Foster}},
  \bibinfo {author} {\bibfnamefont {M.}~\bibnamefont {Wilde}}, \bibinfo
  {author} {\bibfnamefont {K.}~\bibnamefont {Chard}}, \bibinfo {author}
  {\bibfnamefont {A.}~\bibnamefont {Woodard}}, \bibinfo {author} {\bibfnamefont
  {Z.}~\bibnamefont {Li}}, \bibinfo {author} {\bibfnamefont {D.~S.}\
  \bibnamefont {Katz}}, \bibinfo {author} {\bibfnamefont {B.}~\bibnamefont
  {Clifford}}, \bibinfo {author} {\bibfnamefont {R.}~\bibnamefont {Kumar}},
  \bibinfo {author} {\bibfnamefont {L.}~\bibnamefont {Lacinski}}, \bibinfo
  {author} {\bibfnamefont {R.}~\bibnamefont {Chard}}, \ and\ \bibinfo {author}
  {\bibfnamefont {J.~M.}\ \bibnamefont {Wozniak}},\ }in\ \href {\doibase
  10.1145/3307681.3325400} {\emph {\bibinfo {booktitle} {Proceedings of the
  28th International Symposium on High-Performance Parallel and Distributed
  Computing - {HPDC} 2019}}}\ (\bibinfo  {publisher} {{ACM} Press},\ \bibinfo
  {year} {2019})\BibitemShut {NoStop}%
\bibitem [{\citenamefont {Mounet}\ \emph
  {et~al.}(2018{\natexlab{a}})\citenamefont {Mounet}, \citenamefont
  {Gibertini}, \citenamefont {Schwaller}, \citenamefont {Campi}, \citenamefont
  {Merkys}, \citenamefont {Marrazzo}, \citenamefont {Sohier}, \citenamefont
  {Castelli}, \citenamefont {Cepellotti}, \citenamefont {Pizzi},\ and\
  \citenamefont {Marzari}}]{Mounet:2018}%
  \BibitemOpen
  \bibfield  {author} {\bibinfo {author} {\bibfnamefont {N.}~\bibnamefont
  {Mounet}}, \bibinfo {author} {\bibfnamefont {M.}~\bibnamefont {Gibertini}},
  \bibinfo {author} {\bibfnamefont {P.}~\bibnamefont {Schwaller}}, \bibinfo
  {author} {\bibfnamefont {D.}~\bibnamefont {Campi}}, \bibinfo {author}
  {\bibfnamefont {A.}~\bibnamefont {Merkys}}, \bibinfo {author} {\bibfnamefont
  {A.}~\bibnamefont {Marrazzo}}, \bibinfo {author} {\bibfnamefont
  {T.}~\bibnamefont {Sohier}}, \bibinfo {author} {\bibfnamefont {I.~E.}\
  \bibnamefont {Castelli}}, \bibinfo {author} {\bibfnamefont {A.}~\bibnamefont
  {Cepellotti}}, \bibinfo {author} {\bibfnamefont {G.}~\bibnamefont {Pizzi}}, \
  and\ \bibinfo {author} {\bibfnamefont {N.}~\bibnamefont {Marzari}},\ }\href
  {\doibase 10.1038/s41565-017-0035-5} {\bibfield  {journal} {\bibinfo
  {journal} {Nature Nanotechnology}\ }\textbf {\bibinfo {volume} {13}},\
  \bibinfo {pages} {246} (\bibinfo {year} {2018}{\natexlab{a}})}\BibitemShut
  {NoStop}%
\bibitem [{\citenamefont {Kahle}\ \emph {et~al.}(2020)\citenamefont {Kahle},
  \citenamefont {Marcolongo},\ and\ \citenamefont {Marzari}}]{Kahle:2020}%
  \BibitemOpen
  \bibfield  {author} {\bibinfo {author} {\bibfnamefont {L.}~\bibnamefont
  {Kahle}}, \bibinfo {author} {\bibfnamefont {A.}~\bibnamefont {Marcolongo}}, \
  and\ \bibinfo {author} {\bibfnamefont {N.}~\bibnamefont {Marzari}},\ }\href
  {\doibase 10.1039/c9ee02457c} {\bibfield  {journal} {\bibinfo  {journal}
  {Energy {\&} Environmental Science}\ } (\bibinfo {year} {2020}),\
  10.1039/c9ee02457c}\BibitemShut {NoStop}%
\bibitem [{\citenamefont {Mercado}\ \emph {et~al.}(2018)\citenamefont
  {Mercado}, \citenamefont {Fu}, \citenamefont {Yakutovich}, \citenamefont
  {Talirz}, \citenamefont {Haranczyk},\ and\ \citenamefont
  {Smit}}]{Mercado:2018}%
  \BibitemOpen
  \bibfield  {author} {\bibinfo {author} {\bibfnamefont {R.}~\bibnamefont
  {Mercado}}, \bibinfo {author} {\bibfnamefont {R.-S.}\ \bibnamefont {Fu}},
  \bibinfo {author} {\bibfnamefont {A.~V.}\ \bibnamefont {Yakutovich}},
  \bibinfo {author} {\bibfnamefont {L.}~\bibnamefont {Talirz}}, \bibinfo
  {author} {\bibfnamefont {M.}~\bibnamefont {Haranczyk}}, \ and\ \bibinfo
  {author} {\bibfnamefont {B.}~\bibnamefont {Smit}},\ }\href {\doibase
  10.1021/acs.chemmater.8b01425} {\bibfield  {journal} {\bibinfo  {journal}
  {Chemistry of Materials}\ }\textbf {\bibinfo {volume} {30}},\ \bibinfo
  {pages} {5069} (\bibinfo {year} {2018})}\BibitemShut {NoStop}%
\bibitem [{\citenamefont {Prandini}\ \emph {et~al.}(2018)\citenamefont
  {Prandini}, \citenamefont {Marrazzo}, \citenamefont {Castelli}, \citenamefont
  {Mounet},\ and\ \citenamefont {Marzari}}]{Prandini:2018}%
  \BibitemOpen
  \bibfield  {author} {\bibinfo {author} {\bibfnamefont {G.}~\bibnamefont
  {Prandini}}, \bibinfo {author} {\bibfnamefont {A.}~\bibnamefont {Marrazzo}},
  \bibinfo {author} {\bibfnamefont {I.~E.}\ \bibnamefont {Castelli}}, \bibinfo
  {author} {\bibfnamefont {N.}~\bibnamefont {Mounet}}, \ and\ \bibinfo {author}
  {\bibfnamefont {N.}~\bibnamefont {Marzari}},\ }\href {\doibase
  10.1038/s41524-018-0127-2} {\bibfield  {journal} {\bibinfo  {journal} {npj
  Computational Materials}\ }\textbf {\bibinfo {volume} {4}} (\bibinfo {year}
  {2018}),\ 10.1038/s41524-018-0127-2}\BibitemShut {NoStop}%
\bibitem [{\citenamefont {Vitale}\ \emph {et~al.}(2019)\citenamefont {Vitale},
  \citenamefont {Pizzi}, \citenamefont {Marrazzo}, \citenamefont {Yates},
  \citenamefont {Marzari},\ and\ \citenamefont {Mostofi}}]{Vitale:2019}%
  \BibitemOpen
  \bibfield  {author} {\bibinfo {author} {\bibfnamefont {V.}~\bibnamefont
  {Vitale}}, \bibinfo {author} {\bibfnamefont {G.}~\bibnamefont {Pizzi}},
  \bibinfo {author} {\bibfnamefont {A.}~\bibnamefont {Marrazzo}}, \bibinfo
  {author} {\bibfnamefont {J.}~\bibnamefont {Yates}}, \bibinfo {author}
  {\bibfnamefont {N.}~\bibnamefont {Marzari}}, \ and\ \bibinfo {author}
  {\bibfnamefont {A.}~\bibnamefont {Mostofi}},\ }\href@noop {} {\  (\bibinfo
  {year} {2019})},\ \Eprint {http://arxiv.org/abs/1909.00433} {arXiv:1909.00433
  [physics.comp-ph]} \BibitemShut {NoStop}%
\bibitem [{\citenamefont {Talirz}\ \emph {et~al.}(2020)\citenamefont {Talirz},
  \citenamefont {Kumbhar}, \citenamefont {Passaro}, \citenamefont {Yakutovich},
  \citenamefont {Granata}, \citenamefont {Gargiulo}, \citenamefont {Borelli},
  \citenamefont {Uhrin}, \citenamefont {Huber}, \citenamefont {Zoupanos},
  \citenamefont {Adorf}, \citenamefont {Andersen}, \citenamefont {Sch\"utt},
  \citenamefont {Pignedoli}, \citenamefont {Passerone}, \citenamefont
  {VandeVondele}, \citenamefont {Schulthess}, \citenamefont {Smit},
  \citenamefont {Pizzi},\ and\ \citenamefont {Marzari}}]{Talirz:2020}%
  \BibitemOpen
  \bibfield  {author} {\bibinfo {author} {\bibfnamefont {L.}~\bibnamefont
  {Talirz}}, \bibinfo {author} {\bibfnamefont {S.}~\bibnamefont {Kumbhar}},
  \bibinfo {author} {\bibfnamefont {E.}~\bibnamefont {Passaro}}, \bibinfo
  {author} {\bibfnamefont {A.~V.}\ \bibnamefont {Yakutovich}}, \bibinfo
  {author} {\bibfnamefont {V.}~\bibnamefont {Granata}}, \bibinfo {author}
  {\bibfnamefont {F.}~\bibnamefont {Gargiulo}}, \bibinfo {author}
  {\bibfnamefont {M.}~\bibnamefont {Borelli}}, \bibinfo {author} {\bibfnamefont
  {M.}~\bibnamefont {Uhrin}}, \bibinfo {author} {\bibfnamefont {S.~P.}\
  \bibnamefont {Huber}}, \bibinfo {author} {\bibfnamefont {S.}~\bibnamefont
  {Zoupanos}}, \bibinfo {author} {\bibfnamefont {C.~S.}\ \bibnamefont {Adorf}},
  \bibinfo {author} {\bibfnamefont {C.~W.}\ \bibnamefont {Andersen}}, \bibinfo
  {author} {\bibfnamefont {O.}~\bibnamefont {Sch\"utt}}, \bibinfo {author}
  {\bibfnamefont {C.~A.}\ \bibnamefont {Pignedoli}}, \bibinfo {author}
  {\bibfnamefont {D.}~\bibnamefont {Passerone}}, \bibinfo {author}
  {\bibfnamefont {J.}~\bibnamefont {VandeVondele}}, \bibinfo {author}
  {\bibfnamefont {T.~C.}\ \bibnamefont {Schulthess}}, \bibinfo {author}
  {\bibfnamefont {B.}~\bibnamefont {Smit}}, \bibinfo {author} {\bibfnamefont
  {G.}~\bibnamefont {Pizzi}}, \ and\ \bibinfo {author} {\bibfnamefont
  {N.}~\bibnamefont {Marzari}},\ }\href@noop {} {\bibfield  {journal} {\bibinfo
   {journal} {in preparation}\ } (\bibinfo {year} {2020})}\BibitemShut
  {NoStop}%
\bibitem [{\citenamefont {Ullmann}(1976)}]{Ullmann:1976}%
  \BibitemOpen
  \bibfield  {author} {\bibinfo {author} {\bibfnamefont {J.~R.}\ \bibnamefont
  {Ullmann}},\ }\href {\doibase 10.1145/321921.321925} {\bibfield  {journal}
  {\bibinfo  {journal} {Journal of the {ACM} ({JACM})}\ }\textbf {\bibinfo
  {volume} {23}},\ \bibinfo {pages} {31} (\bibinfo {year} {1976})}\BibitemShut
  {NoStop}%
\bibitem [{\citenamefont {Curtarolo}\ \emph {et~al.}(2012)\citenamefont
  {Curtarolo}, \citenamefont {Setyawan}, \citenamefont {Hart}, \citenamefont
  {Jahnatek}, \citenamefont {Chepulskii}, \citenamefont {Taylor}, \citenamefont
  {Wang}, \citenamefont {Xue}, \citenamefont {Yang}, \citenamefont {Levy},
  \citenamefont {Mehl}, \citenamefont {Stokes}, \citenamefont {Demchenko},\
  and\ \citenamefont {Morgan}}]{Curtarolo:2012}%
  \BibitemOpen
  \bibfield  {author} {\bibinfo {author} {\bibfnamefont {S.}~\bibnamefont
  {Curtarolo}}, \bibinfo {author} {\bibfnamefont {W.}~\bibnamefont {Setyawan}},
  \bibinfo {author} {\bibfnamefont {G.~L.}\ \bibnamefont {Hart}}, \bibinfo
  {author} {\bibfnamefont {M.}~\bibnamefont {Jahnatek}}, \bibinfo {author}
  {\bibfnamefont {R.~V.}\ \bibnamefont {Chepulskii}}, \bibinfo {author}
  {\bibfnamefont {R.~H.}\ \bibnamefont {Taylor}}, \bibinfo {author}
  {\bibfnamefont {S.}~\bibnamefont {Wang}}, \bibinfo {author} {\bibfnamefont
  {J.}~\bibnamefont {Xue}}, \bibinfo {author} {\bibfnamefont {K.}~\bibnamefont
  {Yang}}, \bibinfo {author} {\bibfnamefont {O.}~\bibnamefont {Levy}}, \bibinfo
  {author} {\bibfnamefont {M.~J.}\ \bibnamefont {Mehl}}, \bibinfo {author}
  {\bibfnamefont {H.~T.}\ \bibnamefont {Stokes}}, \bibinfo {author}
  {\bibfnamefont {D.~O.}\ \bibnamefont {Demchenko}}, \ and\ \bibinfo {author}
  {\bibfnamefont {D.}~\bibnamefont {Morgan}},\ }\href {\doibase
  10.1016/j.commatsci.2012.02.005} {\bibfield  {journal} {\bibinfo  {journal}
  {Computational Materials Science}\ }\textbf {\bibinfo {volume} {58}},\
  \bibinfo {pages} {218} (\bibinfo {year} {2012})}\BibitemShut {NoStop}%
\bibitem [{\citenamefont {Gra{\v{z}}ulis}\ \emph {et~al.}(2011)\citenamefont
  {Gra{\v{z}}ulis}, \citenamefont {Da{\v{s}}kevi{\v{c}}}, \citenamefont
  {Merkys}, \citenamefont {Chateigner}, \citenamefont {Lutterotti},
  \citenamefont {Quir{\'{o}}s}, \citenamefont {Serebryanaya}, \citenamefont
  {Moeck}, \citenamefont {Downs},\ and\ \citenamefont {Bail}}]{Grazulis:2012}%
  \BibitemOpen
  \bibfield  {author} {\bibinfo {author} {\bibfnamefont {S.}~\bibnamefont
  {Gra{\v{z}}ulis}}, \bibinfo {author} {\bibfnamefont {A.}~\bibnamefont
  {Da{\v{s}}kevi{\v{c}}}}, \bibinfo {author} {\bibfnamefont {A.}~\bibnamefont
  {Merkys}}, \bibinfo {author} {\bibfnamefont {D.}~\bibnamefont {Chateigner}},
  \bibinfo {author} {\bibfnamefont {L.}~\bibnamefont {Lutterotti}}, \bibinfo
  {author} {\bibfnamefont {M.}~\bibnamefont {Quir{\'{o}}s}}, \bibinfo {author}
  {\bibfnamefont {N.~R.}\ \bibnamefont {Serebryanaya}}, \bibinfo {author}
  {\bibfnamefont {P.}~\bibnamefont {Moeck}}, \bibinfo {author} {\bibfnamefont
  {R.~T.}\ \bibnamefont {Downs}}, \ and\ \bibinfo {author} {\bibfnamefont
  {A.~L.}\ \bibnamefont {Bail}},\ }\href {\doibase 10.1093/nar/gkr900}
  {\bibfield  {journal} {\bibinfo  {journal} {Nucleic Acids Research}\ }\textbf
  {\bibinfo {volume} {40}},\ \bibinfo {pages} {D420} (\bibinfo {year}
  {2011})}\BibitemShut {NoStop}%
\bibitem [{\citenamefont {Jain}\ \emph {et~al.}(2013)\citenamefont {Jain},
  \citenamefont {Ong}, \citenamefont {Hautier}, \citenamefont {Chen},
  \citenamefont {Richards}, \citenamefont {Dacek}, \citenamefont {Cholia},
  \citenamefont {Gunter}, \citenamefont {Skinner}, \citenamefont {Ceder},\ and\
  \citenamefont {Persson}}]{Jain:2013}%
  \BibitemOpen
  \bibfield  {author} {\bibinfo {author} {\bibfnamefont {A.}~\bibnamefont
  {Jain}}, \bibinfo {author} {\bibfnamefont {S.~P.}\ \bibnamefont {Ong}},
  \bibinfo {author} {\bibfnamefont {G.}~\bibnamefont {Hautier}}, \bibinfo
  {author} {\bibfnamefont {W.}~\bibnamefont {Chen}}, \bibinfo {author}
  {\bibfnamefont {W.~D.}\ \bibnamefont {Richards}}, \bibinfo {author}
  {\bibfnamefont {S.}~\bibnamefont {Dacek}}, \bibinfo {author} {\bibfnamefont
  {S.}~\bibnamefont {Cholia}}, \bibinfo {author} {\bibfnamefont
  {D.}~\bibnamefont {Gunter}}, \bibinfo {author} {\bibfnamefont
  {D.}~\bibnamefont {Skinner}}, \bibinfo {author} {\bibfnamefont
  {G.}~\bibnamefont {Ceder}}, \ and\ \bibinfo {author} {\bibfnamefont {K.~A.}\
  \bibnamefont {Persson}},\ }\href {\doibase 10.1063/1.4812323} {\bibfield
  {journal} {\bibinfo  {journal} {{APL} Materials}\ }\textbf {\bibinfo {volume}
  {1}},\ \bibinfo {pages} {011002} (\bibinfo {year} {2013})}\BibitemShut
  {NoStop}%
\bibitem [{\citenamefont {Kirklin}\ \emph {et~al.}(2015)\citenamefont
  {Kirklin}, \citenamefont {Saal}, \citenamefont {Meredig}, \citenamefont
  {Thompson}, \citenamefont {Doak}, \citenamefont {Aykol}, \citenamefont
  {Rühl},\ and\ \citenamefont {Wolverton}}]{Kirklin:2015}%
  \BibitemOpen
  \bibfield  {author} {\bibinfo {author} {\bibfnamefont {S.}~\bibnamefont
  {Kirklin}}, \bibinfo {author} {\bibfnamefont {J.~E.}\ \bibnamefont {Saal}},
  \bibinfo {author} {\bibfnamefont {B.}~\bibnamefont {Meredig}}, \bibinfo
  {author} {\bibfnamefont {A.}~\bibnamefont {Thompson}}, \bibinfo {author}
  {\bibfnamefont {J.~W.}\ \bibnamefont {Doak}}, \bibinfo {author}
  {\bibfnamefont {M.}~\bibnamefont {Aykol}}, \bibinfo {author} {\bibfnamefont
  {S.}~\bibnamefont {Rühl}}, \ and\ \bibinfo {author} {\bibfnamefont
  {C.}~\bibnamefont {Wolverton}},\ }\href {\doibase
  10.1038/npjcompumats.2015.10} {\bibfield  {journal} {\bibinfo  {journal} {npj
  Computational Materials}\ }\textbf {\bibinfo {volume} {1}} (\bibinfo {year}
  {2015}),\ 10.1038/npjcompumats.2015.10}\BibitemShut {NoStop}%
\bibitem [{\citenamefont {Duvall}\ \emph {et~al.}(2007)\citenamefont {Duvall},
  \citenamefont {Matyas},\ and\ \citenamefont {Glover}}]{Duvall:2007}%
  \BibitemOpen
  \bibfield  {author} {\bibinfo {author} {\bibfnamefont {P.}~\bibnamefont
  {Duvall}}, \bibinfo {author} {\bibfnamefont {S.~M.}\ \bibnamefont {Matyas}},
  \ and\ \bibinfo {author} {\bibfnamefont {A.}~\bibnamefont {Glover}},\
  }\href@noop {} {\emph {\bibinfo {title} {Continuous Integration: Improving
  Software Quality and Reducing Risk (The Addison-Wesley Signature Series)}}}\
  (\bibinfo  {publisher} {Addison-Wesley Professional},\ \bibinfo {year}
  {2007})\BibitemShut {NoStop}%
\bibitem [{\citenamefont {Larsen}\ \emph {et~al.}(2017)\citenamefont {Larsen},
  \citenamefont {Mortensen}, \citenamefont {Blomqvist}, \citenamefont
  {Castelli}, \citenamefont {Christensen}, \citenamefont {Du{\l}ak},
  \citenamefont {Friis}, \citenamefont {Groves}, \citenamefont {Hammer},
  \citenamefont {Hargus}, \citenamefont {Hermes}, \citenamefont {Jennings},
  \citenamefont {Jensen}, \citenamefont {Kermode}, \citenamefont {Kitchin},
  \citenamefont {Kolsbjerg}, \citenamefont {Kubal}, \citenamefont {Kaasbjerg},
  \citenamefont {Lysgaard}, \citenamefont {Maronsson}, \citenamefont {Maxson},
  \citenamefont {Olsen}, \citenamefont {Pastewka}, \citenamefont {Peterson},
  \citenamefont {Rostgaard}, \citenamefont {Schi{\o}tz}, \citenamefont
  {Sch\"utt}, \citenamefont {Strange}, \citenamefont {Thygesen}, \citenamefont
  {Vegge}, \citenamefont {Vilhelmsen}, \citenamefont {Walter}, \citenamefont
  {Zeng},\ and\ \citenamefont {Jacobsen}}]{HjorthLarsen:2017}%
  \BibitemOpen
  \bibfield  {author} {\bibinfo {author} {\bibfnamefont {A.~H.}\ \bibnamefont
  {Larsen}}, \bibinfo {author} {\bibfnamefont {J.~J.}\ \bibnamefont
  {Mortensen}}, \bibinfo {author} {\bibfnamefont {J.}~\bibnamefont
  {Blomqvist}}, \bibinfo {author} {\bibfnamefont {I.~E.}\ \bibnamefont
  {Castelli}}, \bibinfo {author} {\bibfnamefont {R.}~\bibnamefont
  {Christensen}}, \bibinfo {author} {\bibfnamefont {M.}~\bibnamefont
  {Du{\l}ak}}, \bibinfo {author} {\bibfnamefont {J.}~\bibnamefont {Friis}},
  \bibinfo {author} {\bibfnamefont {M.~N.}\ \bibnamefont {Groves}}, \bibinfo
  {author} {\bibfnamefont {B.}~\bibnamefont {Hammer}}, \bibinfo {author}
  {\bibfnamefont {C.}~\bibnamefont {Hargus}}, \bibinfo {author} {\bibfnamefont
  {E.~D.}\ \bibnamefont {Hermes}}, \bibinfo {author} {\bibfnamefont {P.~C.}\
  \bibnamefont {Jennings}}, \bibinfo {author} {\bibfnamefont {P.~B.}\
  \bibnamefont {Jensen}}, \bibinfo {author} {\bibfnamefont {J.}~\bibnamefont
  {Kermode}}, \bibinfo {author} {\bibfnamefont {J.~R.}\ \bibnamefont
  {Kitchin}}, \bibinfo {author} {\bibfnamefont {E.~L.}\ \bibnamefont
  {Kolsbjerg}}, \bibinfo {author} {\bibfnamefont {J.}~\bibnamefont {Kubal}},
  \bibinfo {author} {\bibfnamefont {K.}~\bibnamefont {Kaasbjerg}}, \bibinfo
  {author} {\bibfnamefont {S.}~\bibnamefont {Lysgaard}}, \bibinfo {author}
  {\bibfnamefont {J.~B.}\ \bibnamefont {Maronsson}}, \bibinfo {author}
  {\bibfnamefont {T.}~\bibnamefont {Maxson}}, \bibinfo {author} {\bibfnamefont
  {T.}~\bibnamefont {Olsen}}, \bibinfo {author} {\bibfnamefont
  {L.}~\bibnamefont {Pastewka}}, \bibinfo {author} {\bibfnamefont
  {A.}~\bibnamefont {Peterson}}, \bibinfo {author} {\bibfnamefont
  {C.}~\bibnamefont {Rostgaard}}, \bibinfo {author} {\bibfnamefont
  {J.}~\bibnamefont {Schi{\o}tz}}, \bibinfo {author} {\bibfnamefont
  {O.}~\bibnamefont {Sch\"utt}}, \bibinfo {author} {\bibfnamefont
  {M.}~\bibnamefont {Strange}}, \bibinfo {author} {\bibfnamefont {K.~S.}\
  \bibnamefont {Thygesen}}, \bibinfo {author} {\bibfnamefont {T.}~\bibnamefont
  {Vegge}}, \bibinfo {author} {\bibfnamefont {L.}~\bibnamefont {Vilhelmsen}},
  \bibinfo {author} {\bibfnamefont {M.}~\bibnamefont {Walter}}, \bibinfo
  {author} {\bibfnamefont {Z.}~\bibnamefont {Zeng}}, \ and\ \bibinfo {author}
  {\bibfnamefont {K.~W.}\ \bibnamefont {Jacobsen}},\ }\href {\doibase
  10.1088/1361-648x/aa680e} {\bibfield  {journal} {\bibinfo  {journal} {Journal
  of Physics: Condensed Matter}\ }\textbf {\bibinfo {volume} {29}},\ \bibinfo
  {pages} {273002} (\bibinfo {year} {2017})}\BibitemShut {NoStop}%
\bibitem [{\citenamefont {Ong}\ \emph {et~al.}(2013)\citenamefont {Ong},
  \citenamefont {Richards}, \citenamefont {Jain}, \citenamefont {Hautier},
  \citenamefont {Kocher}, \citenamefont {Cholia}, \citenamefont {Gunter},
  \citenamefont {Chevrier}, \citenamefont {Persson},\ and\ \citenamefont
  {Ceder}}]{Ong:2013}%
  \BibitemOpen
  \bibfield  {author} {\bibinfo {author} {\bibfnamefont {S.~P.}\ \bibnamefont
  {Ong}}, \bibinfo {author} {\bibfnamefont {W.~D.}\ \bibnamefont {Richards}},
  \bibinfo {author} {\bibfnamefont {A.}~\bibnamefont {Jain}}, \bibinfo {author}
  {\bibfnamefont {G.}~\bibnamefont {Hautier}}, \bibinfo {author} {\bibfnamefont
  {M.}~\bibnamefont {Kocher}}, \bibinfo {author} {\bibfnamefont
  {S.}~\bibnamefont {Cholia}}, \bibinfo {author} {\bibfnamefont
  {D.}~\bibnamefont {Gunter}}, \bibinfo {author} {\bibfnamefont {V.~L.}\
  \bibnamefont {Chevrier}}, \bibinfo {author} {\bibfnamefont {K.~A.}\
  \bibnamefont {Persson}}, \ and\ \bibinfo {author} {\bibfnamefont
  {G.}~\bibnamefont {Ceder}},\ }\href {\doibase
  10.1016/j.commatsci.2012.10.028} {\bibfield  {journal} {\bibinfo  {journal}
  {Computational Materials Science}\ }\textbf {\bibinfo {volume} {68}},\
  \bibinfo {pages} {314} (\bibinfo {year} {2013})}\BibitemShut {NoStop}%
\bibitem [{\citenamefont {Togo}\ and\ \citenamefont
  {Tanaka}(2018)}]{Atsushi:2018}%
  \BibitemOpen
  \bibfield  {author} {\bibinfo {author} {\bibfnamefont {A.}~\bibnamefont
  {Togo}}\ and\ \bibinfo {author} {\bibfnamefont {I.}~\bibnamefont {Tanaka}},\
  }\href {https://arxiv.org/abs/1808.01590} {\bibfield  {journal} {\bibinfo
  {journal} {ArXiv e-prints}\ } (\bibinfo {year} {2018})},\ \Eprint
  {http://arxiv.org/abs/1808.01590} {1808.01590} \BibitemShut {NoStop}%
\bibitem [{\citenamefont {Hinuma}\ \emph {et~al.}(2017)\citenamefont {Hinuma},
  \citenamefont {Pizzi}, \citenamefont {Kumagai}, \citenamefont {Oba},\ and\
  \citenamefont {Tanaka}}]{Hinuma:2017}%
  \BibitemOpen
  \bibfield  {author} {\bibinfo {author} {\bibfnamefont {Y.}~\bibnamefont
  {Hinuma}}, \bibinfo {author} {\bibfnamefont {G.}~\bibnamefont {Pizzi}},
  \bibinfo {author} {\bibfnamefont {Y.}~\bibnamefont {Kumagai}}, \bibinfo
  {author} {\bibfnamefont {F.}~\bibnamefont {Oba}}, \ and\ \bibinfo {author}
  {\bibfnamefont {I.}~\bibnamefont {Tanaka}},\ }\href {\doibase
  10.1016/j.commatsci.2016.10.015} {\bibfield  {journal} {\bibinfo  {journal}
  {Computational Materials Science}\ }\textbf {\bibinfo {volume} {128}},\
  \bibinfo {pages} {140} (\bibinfo {year} {2017})}\BibitemShut {NoStop}%
\bibitem [{\citenamefont {Belsky}\ \emph {et~al.}(2002)\citenamefont {Belsky},
  \citenamefont {Hellenbrandt}, \citenamefont {Karen},\ and\ \citenamefont
  {Luksch}}]{Belsky:2002}%
  \BibitemOpen
  \bibfield  {author} {\bibinfo {author} {\bibfnamefont {A.}~\bibnamefont
  {Belsky}}, \bibinfo {author} {\bibfnamefont {M.}~\bibnamefont
  {Hellenbrandt}}, \bibinfo {author} {\bibfnamefont {V.~L.}\ \bibnamefont
  {Karen}}, \ and\ \bibinfo {author} {\bibfnamefont {P.}~\bibnamefont
  {Luksch}},\ }\href {\doibase 10.1107/s0108768102006948} {\bibfield  {journal}
  {\bibinfo  {journal} {Acta Crystallographica Section B Structural Science}\
  }\textbf {\bibinfo {volume} {58}},\ \bibinfo {pages} {364} (\bibinfo {year}
  {2002})}\BibitemShut {NoStop}%
\bibitem [{\citenamefont {Merkys}\ \emph {et~al.}(2017)\citenamefont {Merkys},
  \citenamefont {Mounet}, \citenamefont {Cepellotti}, \citenamefont {Marzari},
  \citenamefont {Gra{\v{z}}ulis},\ and\ \citenamefont {Pizzi}}]{Merkys:2017}%
  \BibitemOpen
  \bibfield  {author} {\bibinfo {author} {\bibfnamefont {A.}~\bibnamefont
  {Merkys}}, \bibinfo {author} {\bibfnamefont {N.}~\bibnamefont {Mounet}},
  \bibinfo {author} {\bibfnamefont {A.}~\bibnamefont {Cepellotti}}, \bibinfo
  {author} {\bibfnamefont {N.}~\bibnamefont {Marzari}}, \bibinfo {author}
  {\bibfnamefont {S.}~\bibnamefont {Gra{\v{z}}ulis}}, \ and\ \bibinfo {author}
  {\bibfnamefont {G.}~\bibnamefont {Pizzi}},\ }\href {\doibase
  10.1186/s13321-017-0242-y} {\bibfield  {journal} {\bibinfo  {journal}
  {Journal of Cheminformatics}\ }\textbf {\bibinfo {volume} {9}} (\bibinfo
  {year} {2017}),\ 10.1186/s13321-017-0242-y}\BibitemShut {NoStop}%
\bibitem [{\citenamefont {Mounet}\ \emph
  {et~al.}(2018{\natexlab{b}})\citenamefont {Mounet}, \citenamefont
  {Gibertini}, \citenamefont {Schwaller}, \citenamefont {Campi}, \citenamefont
  {Merkys}, \citenamefont {Marrazzo}, \citenamefont {Sohier}, \citenamefont
  {Castelli}, \citenamefont {Cepellotti}, \citenamefont {Pizzi},\ and\
  \citenamefont {Marzari}}]{web:Mounet:MCA:2018}%
  \BibitemOpen
  \bibfield  {author} {\bibinfo {author} {\bibfnamefont {N.}~\bibnamefont
  {Mounet}}, \bibinfo {author} {\bibfnamefont {M.}~\bibnamefont {Gibertini}},
  \bibinfo {author} {\bibfnamefont {P.}~\bibnamefont {Schwaller}}, \bibinfo
  {author} {\bibfnamefont {D.}~\bibnamefont {Campi}}, \bibinfo {author}
  {\bibfnamefont {A.}~\bibnamefont {Merkys}}, \bibinfo {author} {\bibfnamefont
  {A.}~\bibnamefont {Marrazzo}}, \bibinfo {author} {\bibfnamefont
  {T.}~\bibnamefont {Sohier}}, \bibinfo {author} {\bibfnamefont {I.~E.}\
  \bibnamefont {Castelli}}, \bibinfo {author} {\bibfnamefont {A.}~\bibnamefont
  {Cepellotti}}, \bibinfo {author} {\bibfnamefont {G.}~\bibnamefont {Pizzi}}, \
  and\ \bibinfo {author} {\bibfnamefont {N.}~\bibnamefont {Marzari}},\ }\href
  {\doibase 10.24435/materialscloud:2017.0008/v3} {\bibfield  {journal}
  {\bibinfo  {journal} {Materials Cloud}\ }\textbf {\bibinfo {volume}
  {2017.0008/v3}} (\bibinfo {year} {2018}{\natexlab{b}}),\
  10.24435/materialscloud:2017.0008/v3}\BibitemShut {NoStop}%
\bibitem [{\citenamefont {Huber}\ \emph {et~al.}(2020)\citenamefont {Huber},
  \citenamefont {Zoupanos}, \citenamefont {Uhrin}, \citenamefont {Talirz},
  \citenamefont {Kahle}, \citenamefont {H\"auselmann}, \citenamefont {Gresch},
  \citenamefont {M\"uller}, \citenamefont {Yakutovich}, \citenamefont
  {Andersen}, \citenamefont {Ramirez}, \citenamefont {Adorf}, \citenamefont
  {Gargiulo}, \citenamefont {Kumbhar}, \citenamefont {Passaro}, \citenamefont
  {Johnston}, \citenamefont {Merkys}, \citenamefont {Cepellotti}, \citenamefont
  {Mounet}, \citenamefont {Marzari}, \citenamefont {Kozinsky},\ and\
  \citenamefont {Pizzi}}]{web:Huber:MCA:2020}%
  \BibitemOpen
  \bibfield  {author} {\bibinfo {author} {\bibfnamefont {S.~P.}\ \bibnamefont
  {Huber}}, \bibinfo {author} {\bibfnamefont {S.}~\bibnamefont {Zoupanos}},
  \bibinfo {author} {\bibfnamefont {M.}~\bibnamefont {Uhrin}}, \bibinfo
  {author} {\bibfnamefont {L.}~\bibnamefont {Talirz}}, \bibinfo {author}
  {\bibfnamefont {L.}~\bibnamefont {Kahle}}, \bibinfo {author} {\bibfnamefont
  {R.}~\bibnamefont {H\"auselmann}}, \bibinfo {author} {\bibfnamefont
  {D.}~\bibnamefont {Gresch}}, \bibinfo {author} {\bibfnamefont
  {T.}~\bibnamefont {M\"uller}}, \bibinfo {author} {\bibfnamefont {A.~V.}\
  \bibnamefont {Yakutovich}}, \bibinfo {author} {\bibfnamefont {C.~W.}\
  \bibnamefont {Andersen}}, \bibinfo {author} {\bibfnamefont {F.~F.}\
  \bibnamefont {Ramirez}}, \bibinfo {author} {\bibfnamefont {C.~S.}\
  \bibnamefont {Adorf}}, \bibinfo {author} {\bibfnamefont {F.}~\bibnamefont
  {Gargiulo}}, \bibinfo {author} {\bibfnamefont {S.}~\bibnamefont {Kumbhar}},
  \bibinfo {author} {\bibfnamefont {E.}~\bibnamefont {Passaro}}, \bibinfo
  {author} {\bibfnamefont {C.}~\bibnamefont {Johnston}}, \bibinfo {author}
  {\bibfnamefont {A.}~\bibnamefont {Merkys}}, \bibinfo {author} {\bibfnamefont
  {A.}~\bibnamefont {Cepellotti}}, \bibinfo {author} {\bibfnamefont
  {N.}~\bibnamefont {Mounet}}, \bibinfo {author} {\bibfnamefont
  {N.}~\bibnamefont {Marzari}}, \bibinfo {author} {\bibfnamefont
  {B.}~\bibnamefont {Kozinsky}}, \ and\ \bibinfo {author} {\bibfnamefont
  {G.}~\bibnamefont {Pizzi}},\ }\href {\doibase
  10.24435/materialscloud:2020.0027/V1} {\bibfield  {journal} {\bibinfo
  {journal} {Materials Cloud}\ }\textbf {\bibinfo {volume} {2020.0027/v1}}
  (\bibinfo {year} {2020}),\ 10.24435/materialscloud:2020.0027/V1}\BibitemShut
  {NoStop}%
\end{thebibliography}%


\begin{thebibliography}{2}%
\makeatletter
\providecommand \@ifxundefined [1]{%
 \@ifx{#1\undefined}
}%
\providecommand \@ifnum [1]{%
 \ifnum #1\expandafter \@firstoftwo
 \else \expandafter \@secondoftwo
 \fi
}%
\providecommand \@ifx [1]{%
 \ifx #1\expandafter \@firstoftwo
 \else \expandafter \@secondoftwo
 \fi
}%
\providecommand \natexlab [1]{#1}%
\providecommand \enquote  [1]{``#1''}%
\providecommand \bibnamefont  [1]{#1}%
\providecommand \bibfnamefont [1]{#1}%
\providecommand \citenamefont [1]{#1}%
\providecommand \href@noop [0]{\@secondoftwo}%
\providecommand \href [0]{\begingroup \@sanitize@url \@href}%
\providecommand \@href[1]{\@@startlink{#1}\@@href}%
\providecommand \@@href[1]{\endgroup#1\@@endlink}%
\providecommand \@sanitize@url [0]{\catcode `\\12\catcode `\$12\catcode
  `\&12\catcode `\#12\catcode `\^12\catcode `\_12\catcode `\%12\relax}%
\providecommand \@@startlink[1]{}%
\providecommand \@@endlink[0]{}%
\providecommand \url  [0]{\begingroup\@sanitize@url \@url }%
\providecommand \@url [1]{\endgroup\@href {#1}{\urlprefix }}%
\providecommand \urlprefix  [0]{URL }%
\providecommand \Eprint [0]{\href }%
\providecommand \doibase [0]{http://dx.doi.org/}%
\providecommand \selectlanguage [0]{\@gobble}%
\providecommand \bibinfo  [0]{\@secondoftwo}%
\providecommand \bibfield  [0]{\@secondoftwo}%
\providecommand \translation [1]{[#1]}%
\providecommand \BibitemOpen [0]{}%
\providecommand \bibitemStop [0]{}%
\providecommand \bibitemNoStop [0]{.\EOS\space}%
\providecommand \EOS [0]{\spacefactor3000\relax}%
\providecommand \BibitemShut  [1]{\csname bibitem#1\endcsname}%
\let\auto@bib@innerbib\@empty
\bibitem [{\citenamefont {Pizzi}\ \emph {et~al.}(2016)\citenamefont {Pizzi},
  \citenamefont {Cepellotti}, \citenamefont {Sabatini}, \citenamefont
  {Marzari},\ and\ \citenamefont {Kozinsky}}]{Pizzi:2016}%
  \BibitemOpen
  \bibfield  {author} {\bibinfo {author} {\bibfnamefont {G.}~\bibnamefont
  {Pizzi}}, \bibinfo {author} {\bibfnamefont {A.}~\bibnamefont {Cepellotti}},
  \bibinfo {author} {\bibfnamefont {R.}~\bibnamefont {Sabatini}}, \bibinfo
  {author} {\bibfnamefont {N.}~\bibnamefont {Marzari}}, \ and\ \bibinfo
  {author} {\bibfnamefont {B.}~\bibnamefont {Kozinsky}},\ }\href {\doibase
  10.1016/j.commatsci.2015.09.013} {\bibfield  {journal} {\bibinfo  {journal}
  {Computational Materials Science}\ }\textbf {\bibinfo {volume} {111}},\
  \bibinfo {pages} {218} (\bibinfo {year} {2016})}\BibitemShut {NoStop}%
\bibitem [{\citenamefont {Talirz}\ \emph {et~al.}(2020)\citenamefont {Talirz},
  \citenamefont {Kumbhar}, \citenamefont {Passaro}, \citenamefont {Yakutovich},
  \citenamefont {Granata}, \citenamefont {Gargiulo}, \citenamefont {Borelli},
  \citenamefont {Uhrin}, \citenamefont {Huber}, \citenamefont {Zoupanos},
  \citenamefont {Adorf}, \citenamefont {Andersen}, \citenamefont {Sch\"utt},
  \citenamefont {Pignedoli}, \citenamefont {Passerone}, \citenamefont
  {VandeVondele}, \citenamefont {Schulthess}, \citenamefont {Smit},
  \citenamefont {Pizzi},\ and\ \citenamefont {Marzari}}]{Talirz:2020}%
  \BibitemOpen
  \bibfield  {author} {\bibinfo {author} {\bibfnamefont {L.}~\bibnamefont
  {Talirz}}, \bibinfo {author} {\bibfnamefont {S.}~\bibnamefont {Kumbhar}},
  \bibinfo {author} {\bibfnamefont {E.}~\bibnamefont {Passaro}}, \bibinfo
  {author} {\bibfnamefont {A.~V.}\ \bibnamefont {Yakutovich}}, \bibinfo
  {author} {\bibfnamefont {V.}~\bibnamefont {Granata}}, \bibinfo {author}
  {\bibfnamefont {F.}~\bibnamefont {Gargiulo}}, \bibinfo {author}
  {\bibfnamefont {M.}~\bibnamefont {Borelli}}, \bibinfo {author} {\bibfnamefont
  {M.}~\bibnamefont {Uhrin}}, \bibinfo {author} {\bibfnamefont {S.~P.}\
  \bibnamefont {Huber}}, \bibinfo {author} {\bibfnamefont {S.}~\bibnamefont
  {Zoupanos}}, \bibinfo {author} {\bibfnamefont {C.~S.}\ \bibnamefont {Adorf}},
  \bibinfo {author} {\bibfnamefont {C.~W.}\ \bibnamefont {Andersen}}, \bibinfo
  {author} {\bibfnamefont {O.}~\bibnamefont {Sch\"utt}}, \bibinfo {author}
  {\bibfnamefont {C.~A.}\ \bibnamefont {Pignedoli}}, \bibinfo {author}
  {\bibfnamefont {D.}~\bibnamefont {Passerone}}, \bibinfo {author}
  {\bibfnamefont {J.}~\bibnamefont {VandeVondele}}, \bibinfo {author}
  {\bibfnamefont {T.~C.}\ \bibnamefont {Schulthess}}, \bibinfo {author}
  {\bibfnamefont {B.}~\bibnamefont {Smit}}, \bibinfo {author} {\bibfnamefont
  {G.}~\bibnamefont {Pizzi}}, \ and\ \bibinfo {author} {\bibfnamefont
  {N.}~\bibnamefont {Marzari}},\ }\href@noop {} {\bibfield  {journal} {\bibinfo
   {journal} {in preparation}\ } (\bibinfo {year} {2020})}\BibitemShut
  {NoStop}%
\end{thebibliography}%

\end{document}


\title{Supplementary Information: AiiDA 1.0, a scalable computational infrastructure for automated reproducible workflows and data provenance}
\author{}
\maketitle

\subsection{\label{ssec:architecture_differences}Architecture differences with earlier AiiDA versions}
While the ADES model and the overall goals of AiiDA have remained the same as those originally published in Ref.~\cite{Pizzi:2016}, AiiDA now comes with many new features over the 0.x series, with the first public release in Feb 2015 and including 9 major releases and 20 releases in total~(\href{http://www.aiida.net/download/}{aiida.net/download}); most of the code has been completely redesigned.
Improvements aimed towards making AiiDA scalable, and able to support high-throughput computational loads consisting of high-performance computing (HPC) jobs of various size, making it more flexible by simplifying the protocols by which it can be extended, and providing tools and interfaces to facilitate its use.
A schematic overview of the architecture is shown in Fig.~1 of the main paper.

The engine, the component responsible for running all calculations and workflows, is described in detail in ``The engine'' section of the main paper.
In a nutshell, the switch from a polling-based design to an event-based one allows the engine to react instantaneously to state changes, making it much more responsive.
In addition, the new engine is made scalable by allowing an arbitrary number of independent workers to operate in parallel, supporting now sustained throughputs of more than tens of thousands of processes per hour.
Communication between and with the workers is provided by the RabbitMQ message broker~(\href{https://www.rabbitmq.com}{rabbitmq.com}) through the pika client library~(\href{https://pypi.org/project/pika/}{pypi.org/project/pika}).

In addition to increased efficiency, the engine has also been made more robust and now comes with built-in error handling for transient problems, such as connection issues or computational clusters going offline unexpectedly.
Failed executions are automatically rescheduled and, if repeated consecutive failures occur, the calculations are paused such that the problem can be investigated by the user.
The new engine also comes with improvements in terms of usability.
The new workflow language is much more expressive and enables the writing of auto-documenting and reusable workflows.

The two different concepts of calculations and workflows, which behaved and were implemented differently in earlier versions of AiiDA, have now been fully integrated with a homogenised interface, while the entire provenance graph is still stored in PostgreSQL~(\href{http://www.postgresql.org/}{postgresql.org}), a Relational Database Management System (RDMS).
As part of the integration of calculations and workflows, the latter are now also fully part of the provenance graph, with the links between them explicitly represented.
For this reason the ontology of the AiiDA provenance graph has been revisited and rigorously defined as described in ``The provenance model'' section of the main paper.
The extended provenance graph provides more valuable information to the user, while giving them full control over the level of detail with which to inspect it.
To facilitate efficient queries of data and provenance, a dedicated tool has been developed, the \define{QueryBuilder}, allowing users to write advanced graph queries directly in the familiar Python syntax.
The queries are translated to SQL using SQLAlchemy~(\href{http://www.sqlalchemy.org/}{sqlalchemy.org}), a powerful object-relational mapping (ORM) library, which has also been used as a fully-fledged ORM implementation in addition to the original one using Django~(\href{http://www.djangoproject.com}{djangoproject.com}).
To allow different ORMs in AiiDA, the original implementation, initially tightly coupled to Django, was decoupled and an abstract AiiDA frontend ORM was constructed.
This new ORM interface provides a stable API for the user, independent of the backend implementation and makes it easy to implement other backend solutions.

A completely new way of accessing the AiiDA provenance graph is provided by a REST API server (see ``The REST API'' section in the main paper) that allows one to retrieve data over HTTP(S) requests.
This new component enables users to browse the provenance graph programmatically, and implement custom interfaces like the graphical one provided by the Materials Cloud~\cite{Talirz:2020} web platform.

One of the original methods of interacting with AiiDA, the command line interface (CLI) \define{verdi} has been significantly improved.
The management of command-line input has been replaced with the mature \define{Click}~(\href{https://click.palletsprojects.com/en/7.x/}{click.palletsprojects.com}) library.
This guarantees an interface that behaves consistently across all commands and makes it easy to reuse components for additional CLIs that can be distributed through plugins.

Finally, plugins to extend AiiDA's core functionality are now easier to write, share and install thanks to a new flexible plugin system (see ``The plugin system'' section in the main paper).
Although the original version of AiiDA already allowed its functionality to be extended through plugins, the code had to be placed in the source tree of AiiDA itself, tightly coupling the development cycle of the core code and of the plugins.
The new system allows plugins to be developed independently and to be installed with a single command.
A plugin registry~(\href{https://aiidateam.github.io/aiida-registry//}{aiidateam.github.io/aiida-registry/}) has been deployed, serving as a central place where users can discover existing plugins.

The combination of all these changes, which are discussed in detail in the following sections, aim to make AiiDA 1.0 into a powerful and flexible workflow and data management system ready to manage high-throughput HPC jobs on future exascale machines.

\subsection{\label{app:querybuilder}Query builder syntax example}
\begin{figure}
  \centering
  \begin{subfigure}[b]{.48\textwidth}
  \lstset{
    basicstyle=\ttfamily\footnotesize,
    tabsize=1,
    numbers=left,
    upquote=true,
    showstringspaces=false
  }
  \begin{lstlisting}[language=Python]
qb = QueryBuilder()
qb.append(CalcJobNode,
  tag='calculation')
qb.append(Code,
  filters={'label': 'my-code'},
  with_outgoing='calculation')
qb.append(Dict,
  with_outgoing='calculation',
  filters={'attributes.type': 'relax'},
  project=['attributes.threshold'])
qb.append(Dict,
  with_incoming='calculation',
  edge_filters={'label': 'results'},
  project=['attributes.energy'])
  \end{lstlisting}
  \end{subfigure}

  \begin{subfigure}[b]{.40\textwidth}
    \includegraphics[width=\linewidth]{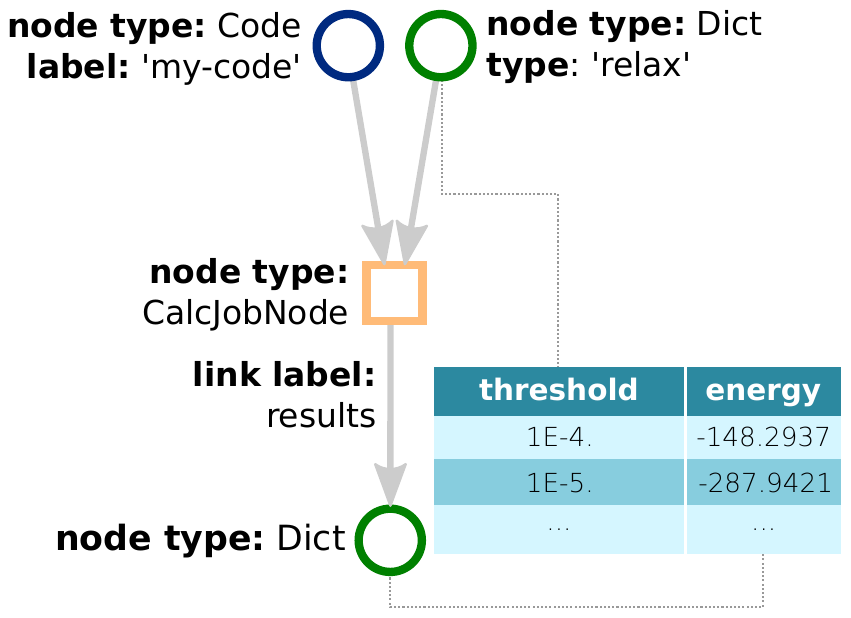}
  \end{subfigure}
  \caption{
    Top: a query builder graph query to filter all calculations that computed the energy of a structure after relaxing it to within a certain threshold, using a code labeled \code{my-code}.
    The result will be a pair of values for each matching result, containing the relaxation threshold (\code{threshold}) and the computed total energy (\code{energy}), retrieved from the attributes of the appropriate nodes.
    The details of the query syntax are explained in the main text.
    Bottom: graphical representation of the same query.}
  \label{fig:qb_query_example}
\end{figure}

To provide a concrete example of the query builder syntax, as implemented in AiiDA's \code{QueryBuilder} class, we show in Fig.~\ref{fig:qb_query_example} a sample query together with a graphical representation of its action and result.
In this example, we query for calculations run with a specific code to compute the total energy of a crystal structure after having relaxed it to within a certain threshold.
The goal is to obtain the total energy of the structure as a function of the relaxation threshold.

To run a new query, first a new query object \code{qb} is created (line 1).
Nodes to be matched are specified by using the \code{append} method of the query object, in which also additional filters can be declared together with the relation to the other nodes in the query.
Finally, ``projections'' indicate which specific properties of the node should be returned as query results.

To perform the query, we first instruct that we are looking for \code{CalcJobNode}s (lines 2-3) that we tag as \code{calculation} to be able to refer to it later when defining inter-node relationships.
The calculation should have a code (line 4) with label \code{my-code} (line 5) as an input (line 6).
Additionally, the calculation should have a \code{Dict} input (line 7-8) containing an attribute \code{type} with value \code{relax} (line 9).
Instead of the entire input dictionary, we request only the value of the \code{threshold} attribute to be returned (line 10).
Finally, the calculation needs to have an output \code{Dict} node (lines 11-14) connected by a link with label \code{results} (line 13), and we project the \code{energy} attribute (line 14).

To evaluate the query, one can then simply run the method \code{qb.all()} that returns a list of results, one for each subgraph matching the query.
Each result is an ordered list of the values that were defined by the projections, in this case the two values (\code{threshold}, \code{energy}).
The final query result then has the form \code{[(threshold1, energy1), (threshold2, energy2), ...]}.

\subsection{Example of a work chain and calculation function}
\begin{figure*}
  \centering
  \begin{subfigure}[b]{.50\linewidth}
  \lstset{
    basicstyle=\ttfamily\footnotesize,
    tabsize=1,
    numbers=left,
    upquote=true,
    showstringspaces=false
  }
\begin{lstlisting}[language=Python]
@calcfunction
def add(x, y):
  return x + y

class Fibonacci(WorkChain):

  @classmethod
  def define(cls, spec):
    super(Fibonacci, cls).define(spec)
    spec.input('N', valid_type=orm.Int,
      help='Compute the Nth Fibonacci number.'
    )
    spec.outline(
      cls.initialize,
      while_(cls.should_iterate)(
        cls.iterate),
      cls.results)
    spec.output('number', valid_type=orm.Int)

  def initialize(self):
    self.ctx.iteration = 0
    self.ctx.previous = orm.Int(0)
    self.ctx.current = orm.Int(1)

  def should_iterate(self):
    return self.ctx.iteration < self.inputs.N - 1

  def iterate(self):
    previous = self.ctx.current
    self.ctx.current = add(
      self.ctx.previous, self.ctx.current)
    self.ctx.previous = previous
    self.ctx.iteration += 1

  def results(self):
    self.out('number', self.ctx.current)

number = run(Fibonacci, N=orm.Int(5))
  \end{lstlisting}
  \end{subfigure}
  \begin{subfigure}[b]{.48\linewidth}
    \centering
    \includegraphics[width=\linewidth]{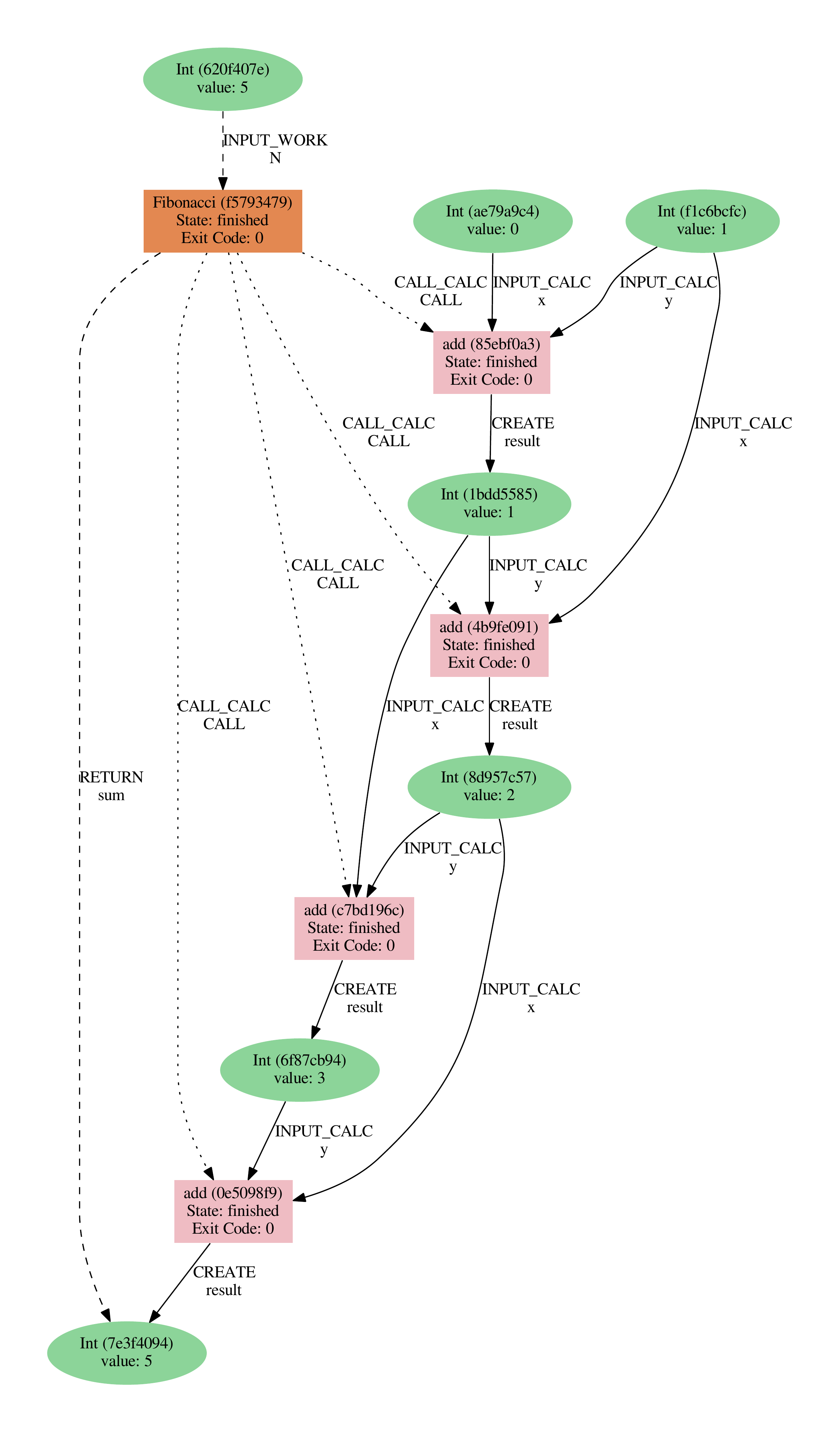}
  \end{subfigure}
  \caption{
    (a) Example implementation to compute the $N^{\mathrm{th}}$ Fibonacci number using AiiDA's workflow engine.
    The \code{Fibonacci} work chain implements Eq.~\eqref{eq:fibonacci} and leverages the \code{add} calculation function to perform the additions.
    (b) Provenance graph of the execution of the \code{Fibonacci} work chain with $N=5$ as input, executed with the last line of panel (a) and returning $f_5=5$ as a result.
  }
  \label{fig:fibonacci}
\end{figure*}

To showcase the interface of AiiDA's workflow language described in the ``The engine'' section of the main paper, we implement a simple Fibonacci number calculator.
The Fibonacci sequence is defined as:
\begin{equation}
  \label{eq:fibonacci}
  f_{N} = f_{N-1} + f_{N-2}
\end{equation}
where $f_0 = 0$ and $f_1 = 1$.
Fig.~\ref{fig:fibonacci}(a) shows a possible implementation using a work chain and a calculation function.
The \code{outline} codifies the logical sequence of the work chain.
The first step \code{initialize} will set some context variables, such as an \code{iteration} counter and the initial Fibonacci numbers $f_0$ and $f_1$, which correspond to \code{previous} and \code{current}, respectively.
The \code{ctx} member of the work chain is a \define{context} that is persisted between the logical steps and can be used to transfer information between them.
The \code{while\_} logical construct is used to tell the workflow to iterate until $N - 1$ iterations have been performed.
Each iteration consists of summing the integers in the \code{previous} and \code{current} variables, which directly corresponds to Eq.~\eqref{eq:fibonacci}.
For this addition the \code{add} calculation function is called such that the provenance is kept.
Finally, after $N$ iterations, the \code{current} value is returned as the requested Fibonacci \code{number} in the \code{results} step.

Fig.~\ref{fig:fibonacci}(b) shows the provenance graph that is produced by AiiDA when running the \code{Fibonacci} work chain.
Note that not only the work chain with the initial input and the final answer is represented, but also all individual additions performed along the way, with their intermediate results.
This implementation of a Fibonacci sequence calculator, while clearly overengineered, shows how arbitrary logic can be implemented in AiiDA's workflow language and how provenance is automatically stored.

\bibliography{manuscript}